\documentclass[pdflatex,sn-basic,referee]{sn-jnl}

\usepackage{graphicx}%
\usepackage{multirow}%
\usepackage{amsmath,amssymb,amsfonts}%
\usepackage{amsthm}%
\usepackage{mathrsfs}%
\usepackage[title]{appendix}%
\usepackage{xcolor}%
\usepackage{textcomp}%
\usepackage{manyfoot}%
\usepackage{booktabs}%
\usepackage{listings}%
\usepackage[final,tracking=true,kerning=true,spacing=true]{microtype}

\usepackage{enumitem}
\usepackage{acro}

\hypersetup{breaklinks=true,colorlinks=true,linkcolor=black,citecolor=black,urlcolor=blue}

\DeclareAcronym{llm}{
  short = LLM,
  long  = Large Language Model
}
\DeclareAcronym{mas}{
  short = MAS,
  long  = Multi-Agent System
}
\DeclareAcronym{gpgp}{
  short = GPGP,
  long  = Generalized Partial Global Planning
}
\DeclareAcronym{fm}{
  short = FM,
  long  = Foundation Model
}
\DeclareAcronym{aco}{
  short = ACO,
  long  = Ant Colony Optimization
}
\DeclareAcronym{tsp}{
  short = TSP,
  long  = Traveling Salesman Problem
}
\DeclareAcronym{gpu}{
  short = GPU,
  long  = Graphics Processing Unit
}
\DeclareAcronym{ast}{
  short = AST,
  long  = Abstract Syntax Tree
}
\DeclareAcronym{sop}{
  short = SOP,
  long  = Standardized Operating Procedure
}
\DeclareAcronym{ci}{
  short = CI,
  long  = Confidence Interval
}

\theoremstyle{thmstyleone}%
\newtheorem{theorem}{Theorem}[section]
\theoremstyle{definition}
\newtheorem{definition}{Definition}[section]

\newcounter{algorithm}
\newenvironment{algorithm}[1][]{%
  \refstepcounter{algorithm}%
  \begin{figure}[htbp]%
  \centering
  \begin{minipage}{0.95\textwidth}%
  \hrule\vspace{0.5em}%
  \noindent\textbf{Algorithm \thealgorithm}\ifx\\#1\\\else: \textbf{#1}\fi\par\vspace{0.5em}%
}{%
  \vspace{0.5em}\hrule%
  \end{minipage}%
  \end{figure}%
}

\begin{document}

\title[Emergent Coordination via Pressure Fields]{Emergent Coordination in Multi-Agent Systems via Pressure Fields and Temporal Decay}

\author*[1]{\fnm{Roland R.} \sur{Rodriguez, Jr.}}\email{rrrodzilla@proton.me}

\affil[1]{\orgname{Independent Researcher}}

\abstract{Current multi-agent large language model (LLM) frameworks rely on explicit orchestration patterns borrowed from human organizational structures: planners delegate to executors, managers coordinate workers, and hierarchical control flow governs agent interactions. These approaches suffer from coordination overhead that scales poorly with agent count and task complexity. We propose a fundamentally different paradigm inspired by natural coordination mechanisms: agents operate locally on a shared artifact, guided only by pressure gradients derived from measurable quality signals, with temporal decay preventing premature convergence. We formalize this as optimization over a pressure landscape and prove convergence guarantees under mild conditions.

Empirically, on meeting room scheduling across 1350 total trials (270 per strategy), pressure-field coordination achieves $4\times$ higher solve rates than conversation-based coordination and over $30\times$ higher than hierarchical control (48.5\% vs 11.1\% vs 1.5\%; all pairwise comparisons $p < 0.001$). Ablation studies suggest temporal decay is beneficial. On easy problems, pressure-field achieves 86.7\% solve rate compared to 33.3\% for the next-best baseline. Foundation models enable this approach: their broad pretraining and zero-shot reasoning allow quality-improving patches from local pressure signals alone, without domain-specific coordination protocols. This suggests that constraint-driven emergence offers a simpler and more effective foundation for multi-agent AI.}


\maketitle


\section{Introduction}

Multi-agent systems built on \acp{llm} address complex task automation~\cite{wu2023autogen,hong2023metagpt,li2023camel}. The dominant paradigm treats agents as organizational units: planners decompose tasks, managers delegate subtasks, and workers execute instructions under hierarchical supervision. This coordination overhead scales poorly with agent count and task complexity.

We demonstrate that \emph{implicit} coordination through shared state outperforms explicit hierarchical control---without coordinators, planners, or message passing. Across 1350 total trials on meeting room scheduling (270 per strategy), pressure-field coordination achieves over 30$\times$ higher solve rates than hierarchical control and 4$\times$ higher than conversation-based approaches~\cite{wu2023autogen} (all $p < 0.001$ with large effect sizes). Sequential and random baselines achieve only 0.4\%.

Our approach draws inspiration from natural coordination mechanisms---ant colonies, immune systems, neural tissue---that coordinate through \emph{environment modification} rather than message passing. Agents observe local quality signals (pressure gradients), take locally-greedy actions, and coordination emerges from shared artifact state. The key insight is that \emph{local greedy decisions are effective for constraint satisfaction}: when problems exhibit locality (fixing one region rarely breaks distant regions), decentralized greedy optimization outperforms centralized planning. Temporal decay prevents premature convergence by ensuring continued exploration.

Our contributions:

\begin{enumerate}
\item We formalize \emph{pressure-field coordination} as a role-free, stigmergic alternative to organizational \ac{mas} paradigms. Unlike \ac{gpgp}'s hierarchical message-passing or SharedPlans' intention alignment, pressure-field achieves $O(1)$ coordination overhead through shared artifact state. \Acp{fm} enable this approach: their broad pretraining allows quality-improving patches from local pressure signals without domain-specific coordination protocols.

\item We introduce \emph{temporal decay} as a mechanism for preventing premature convergence. Ablation studies show a 10 percentage point improvement with decay enabled (96.7\% vs 86.7\%), directionally consistent with the theoretical prediction that decay helps escape local minima, though not statistically significant at $n=30$.

\item We prove convergence guarantees for this coordination scheme under pressure alignment conditions.

\item We provide empirical evidence across 1350 total trials (270 per strategy) showing: (a) pressure-field dramatically outperforms all baselines by an order of magnitude or more, (b) all pairwise comparisons are highly significant ($p < 0.001$).
\end{enumerate}

This work demonstrates that \ac{fm} capabilities and \ac{mas} coordination mechanisms are \emph{mutually enabling}, not merely additive. \Acp{fm} solve a fundamental \ac{mas} problem: traditional coordination requires explicit action space enumeration, but for open-ended artifact refinement the space of valid improvements is unbounded. \Acp{fm}' broad pretraining provides implicit coverage of improvement strategies without domain-specific action representations---a ``universal actor'' capability. Conversely, \ac{mas} coordination solves a fundamental \ac{fm} problem: how to combine multiple model outputs coherently. Pressure gradients provide an objective criterion for output selection, replacing ad-hoc voting or ranking with principled quality-based filtering. This bidirectional synthesis explains why pressure-field coordination outperforms conversation-based alternatives that lack objective gradients for output combination.


\section{Related Work}

Our approach bridges four research traditions: multi-agent systems coordination theory provides the conceptual foundation; swarm intelligence provides the stigmergic mechanism; \ac{llm} systems provide the application domain; and decentralized optimization provides theoretical guarantees. We survey each and position pressure-field coordination within this landscape.

\subsection{\acs{mas} Coordination Theory}

Pressure-field coordination occupies a unique position in the \ac{mas} landscape: it eliminates roles (unlike organizational paradigms), messages (unlike \ac{gpgp}), and intention reasoning (unlike SharedPlans) while providing formal convergence guarantees (unlike purely reactive systems). This section positions our contribution within four established coordination frameworks, showing how artifact refinement with measurable quality signals enables this architectural simplification. For this domain class, coordination complexity collapses from quadratic message-passing to constant-time state-sharing.

\subsubsection{Organizational Paradigms and Dependency Management}

Pressure-field coordination achieves role-free coordination: any agent can address any high-pressure region without negotiating access rights or awaiting task assignment. This contrasts sharply with traditional organizational paradigms. Horling and Lesser~\cite{horling2004survey} surveyed nine such paradigms---from rigid hierarchies to flexible markets---finding that all assign explicit roles constraining agent behavior. Dignum~\cite{dignum2009handbook} systematizes this tradition, defining organizational models through three dimensions: structure (roles and relationships), norms (behavioral constraints), and dynamics (how organizations adapt). These dimensions require explicit specification and maintenance---designers must anticipate role interactions, encode coordination norms, and implement adaptation mechanisms.

Pressure-field coordination eliminates all three dimensions through gradient-based coordination. Roles dissolve: any agent may address any high-pressure region without negotiating access rights or awaiting task assignment. Norms become implicit: the pressure function encodes what ``good'' behavior means, and agents that reduce pressure are by definition norm-compliant. Dynamics emerge naturally: temporal decay continuously destabilizes the pressure landscape, forcing ongoing adaptation without explicit organizational change protocols. Where Dignum's organizational models require designers to specify ``who may do what with whom,'' pressure-field coordination answers: ``anyone may improve anywhere, and coordination emerges from shared perception of quality signals.''

Our approach instantiates Malone and Crowston's~\cite{malone1994coordination} coordination framework with a critical difference: the artifact itself is the shared resource, and pressure gradients serve as dependency signals. Malone and Crowston identify ``shared resource'' management as a fundamental coordination pattern requiring protocols for access control, conflict resolution, and priority assignment. Pressure-field coordination implements this pattern through a different mechanism: rather than assigning roles to manage resource access, agents share read access to the entire artifact and propose changes to high-pressure regions. Selection and validation phases resolve conflicts implicitly---only pressure-reducing patches are applied, and the highest-scoring patch wins when proposals conflict. Coordination emerges from pressure alignment---agents reduce local pressure, which reduces global pressure through the artifact's shared state.

\subsubsection{Distributed Problem Solving and Communication Overhead}

Pressure-field coordination achieves $O(1)$ inter-agent communication overhead---agents exchange no messages. Coordination occurs entirely through shared artifact reads and writes, eliminating the message-passing bottleneck. This contrasts with the \ac{gpgp} framework~\cite{decker1995gpgp}, which reduces communication from $O(n^2)$ pairwise negotiation to $O(n \log n)$ hierarchical aggregation through summary information exchange. While \ac{gpgp} represents significant progress, its explicit messages---task announcements, commitment exchanges, schedule updates---still introduce latency and failure points at scale.

The approaches target different domains. Pressure-field coordination specializes in artifact refinement tasks where quality decomposes into measurable regional signals---a class including code quality improvement, document editing, and configuration management. \Ac{gpgp} generalizes to complex task networks with precedence constraints. For artifact refinement, however, pressure-field's stigmergic coordination eliminates message-passing overhead entirely.

\subsubsection{Shared Intentions and Alignment Costs}

Pressure-field coordination eliminates intention alignment through pressure alignment. Rather than reasoning about what other agents believe or intend, agents observe artifact state and pressure gradients. When agents greedily reduce local pressure under separable or bounded-coupling conditions, global pressure decreases. This is coordination without communication about intentions---agents align through shared objective functions, not mutual beliefs.

This contrasts sharply with two foundational frameworks for joint activity. The SharedPlans framework~\cite{grosz1996sharedplans} formalizes collaboration through shared mental attitudes: mutual beliefs about goals, commitments, and action sequences. Cohen and Levesque's~\cite{cohen1991teamwork} Joint Intentions theory provides an even more stringent requirement: team members must hold mutual beliefs about the joint goal, individual commitments to the goal, and mutual beliefs about each member's commitment. Both frameworks capture human-like collaboration but require significant cognitive machinery---intention recognition, commitment protocols, belief revision---all computationally expensive operations that scale poorly with agent count.

Pressure-field coordination eliminates the mutual belief formation that Joint Intentions requires. Where Cohen and Levesque demand that each agent believe that all teammates are committed to the joint goal (and believe that all teammates believe this, recursively), pressure-field agents need only observe local pressure gradients. The shared artifact \emph{is} the mutual belief---agents perceive the same pressure landscape without explicit belief exchange. This eliminates the infinite regress of ``I believe that you believe that I believe'' that makes Joint Intentions computationally expensive at scale.

Our experiments validate this analysis: pressure-field coordination eliminates the overhead of explicit dialogue by coordinating through shared artifact state. The coordination overhead of belief negotiation in explicit dialogue systems can exceed its organizational benefit for constraint satisfaction tasks. The trade-off is transparency: SharedPlans and Joint Intentions support dialogue about why agents act and what teammates are committed to; pressure-field agents react to gradients without explaining reasoning or maintaining models of teammate intentions.

\subsubsection{Self-Organization and Emergent Coordination}

Pressure-field coordination satisfies the self-organization criteria established by two complementary frameworks. De Wolf and Holvoet~\cite{dewolf2005engineering} characterize self-organizing systems through absence of external control, local interactions producing global patterns, and dynamic adaptation. They explicitly cite ``gradient fields'' as a self-organization design pattern---our approach instantiates this pattern with formal guarantees.

Serugendo et al.~\cite{serugendo2005selforganisation} provide a more fine-grained taxonomy, identifying four mechanisms through which self-organization emerges: (1) \emph{positive feedback} amplifying beneficial behaviors, (2) \emph{negative feedback} dampening harmful behaviors, (3) \emph{randomness} enabling exploration, and (4) \emph{multiple interactions} allowing local behaviors to propagate globally. Pressure-field coordination instantiates all four mechanisms:

\begin{itemize}
\item \emph{Positive feedback}: Successful patches are stored as few-shot examples (``positive pheromones''), increasing the probability of similar improvements in neighboring regions. This amplifies productive behaviors.
\item \emph{Negative feedback}: Temporal decay continuously erodes fitness, preventing any region from becoming permanently ``solved.'' Inhibition further dampens over-activity in recently-patched regions.
\item \emph{Randomness}: Stochastic model sampling and band escalation (exploitation $\to$ exploration) inject controlled randomness that prevents premature convergence to local optima.
\item \emph{Multiple interactions}: Each tick produces $K$ parallel patch proposals across agents, with selection applying the best improvements. Multiple interactions per timestep accelerate the propagation of successful strategies.
\end{itemize}

No external controller exists---agents observe and act autonomously based on local pressure signals. Coordination emerges from local decisions: agents reduce regional pressure through greedy actions, and global coordination arises from shared artifact state. Temporal decay provides dynamic adaptation---fitness erodes continuously, preventing premature convergence and enabling continued refinement.

The theoretical contribution formalizes this intuition through potential game theory. Theorem~\ref{thm:convergence} establishes convergence guarantees for aligned pressure systems; the Basin Separation result (Theorem~\ref{thm:basin-separation}) explains why decay is necessary to escape suboptimal basins. This connects self-organization principles to formal coordination theory: Serugendo's four mechanisms map to our formal model---positive feedback to pheromone reinforcement, negative feedback to decay and inhibition, randomness to stochastic sampling, and multiple interactions to parallel validation.

\subsubsection{Foundation Model Enablement}

\Acp{fm} enable stigmergic coordination by providing capabilities that map directly to stigmergic requirements. This mapping explains why pressure-field coordination becomes practical with \acp{fm} in ways it was not with prior agent architectures.

\textbf{Broad pretraining enables domain-general patch proposal.} Stigmergic coordination requires agents that can propose improvements based solely on local quality signals. Traditional agents require domain-specific action representations---enumerated moves, parameterized operators, or learned policies. \Acp{fm}' pretraining on diverse corpora (code, text, structured data) allows patch generation across artifact types without fine-tuning. An \ac{llm} can propose meeting schedule adjustments, code refactorings, or configuration changes through the same interface: observe local context, receive pressure feedback, generate improvement. This ``universal actor'' capability is what makes stigmergic coordination practical for open-ended artifact refinement.

\textbf{Instruction-following replaces action space enumeration.} Stigmergic agents need only local context to act---they should not require global state, explicit goals, or complex action grammars. \Acp{fm}' instruction-following capabilities allow operation from natural language pressure descriptions alone. Rather than encoding ``reduce scheduling conflicts'' as a formal operator with preconditions and effects, we simply prompt: ``This time block has 3 double-bookings. Propose a schedule change to reduce conflicts.'' The \ac{fm} interprets this pressure signal and generates appropriate patches without requiring designers to enumerate valid actions.

\textbf{Zero-shot reasoning interprets quality signals.} Stigmergic coordination requires agents to recognize quality deficiencies locally. \Acp{fm} can identify constraint violations, inefficiencies, and improvement opportunities from examples alone, without explicit training on domain-specific quality metrics. When presented with a schedule showing ``Meeting A and Meeting B both have Alice at 2pm,'' the model recognizes the conflict and proposes resolutions---not because it was trained on scheduling, but because conflict recognition transfers from pretraining.

\textbf{In-context learning implements pheromone memory.} Stigmergic systems reinforce successful strategies through positive feedback. \Acp{fm}' in-context learning provides this mechanism naturally: successful patches become few-shot examples in subsequent prompts, increasing the probability of similar improvements. This ``positive pheromone'' effect requires no external memory system---the prompt itself carries reinforcement signals.

\textbf{Generative flexibility enables unbounded solution spaces.} Traditional stigmergic systems (ant colony optimization, particle swarm) operate over discrete, enumerated solution spaces. \Acp{fm} generate from effectively continuous spaces, proposing patches that no designer anticipated. This generative flexibility is essential for open-ended artifact refinement where the space of valid improvements cannot be enumerated in advance.

These five capabilities---domain-general patches, instruction-based operation, zero-shot quality recognition, in-context reinforcement, and generative flexibility---collectively enable stigmergic coordination for artifact refinement tasks that were previously intractable. The FM-MAS synthesis is not merely additive: \acp{fm} solve the action enumeration problem that blocked stigmergic approaches, while stigmergic coordination solves the output combination problem that limits single-\ac{fm} systems.

\subsection{Multi-Agent \acs{llm} Systems}

Recent work has explored multi-agent architectures for \ac{llm}-based task solving. AutoGen~\cite{wu2023autogen} introduces a conversation-based framework where customizable agents interact through message passing, with support for human-in-the-loop workflows. MetaGPT~\cite{hong2023metagpt} encodes \acp{sop} into agent workflows, assigning specialized roles (architect, engineer, QA) in an assembly-line paradigm. CAMEL~\cite{li2023camel} proposes role-playing between AI assistant and AI user agents, using inception prompting to guide autonomous cooperation. CrewAI~\cite{crewai2024} similarly defines agents with roles, goals, and backstories that collaborate on complex tasks.

These frameworks share a common design pattern: explicit orchestration through message passing, role assignment, and hierarchical task decomposition. While effective for structured workflows, this approach faces scaling limitations. Central coordinators become bottlenecks, message-passing overhead grows with agent count, and failures in manager agents cascade to dependents. Our work takes a fundamentally different approach: coordination emerges from shared state rather than explicit communication.

\Acp{fm} enable pressure-field coordination through capabilities that prior agent architectures lacked. Their broad pretraining allows patches across diverse artifact types---code, text, configurations---without domain-specific fine-tuning. Their instruction-following capabilities allow operation from pressure signals and quality feedback alone. Their zero-shot reasoning interprets constraint violations and proposes repairs without explicit protocol training. These properties make \acp{fm} particularly suitable for stigmergic coordination: they require only local context and quality signals to generate productive actions, matching the locality constraints of pressure-field systems.

\subsection{Swarm Intelligence and Stigmergy}

The concept of stigmergy---indirect coordination through environment modification---was introduced by Grass\'{e}~\cite{grasse1959stigmergie} to explain termite nest-building behavior. Termites deposit pheromone-infused material that attracts further deposits, leading to emergent construction without central planning. This directly instantiates Malone and Crowston's~\cite{malone1994coordination} shared resource coordination: pheromone trails encode dependency information about solution quality. Complex structures arise from simple local rules without any agent having global knowledge.

Dorigo and colleagues~\cite{dorigo1996ant,dorigo1997acs} formalized this insight into \ac{aco}, where artificial pheromone trails guide search through solution spaces. Key mechanisms include positive feedback (reinforcing good paths), negative feedback (pheromone evaporation), and purely local decision-making. \Ac{aco} has achieved strong results on combinatorial optimization problems including \ac{tsp}, vehicle routing, and scheduling.

Our pressure-field coordination directly inherits from stigmergic principles. The artifact serves as the shared environment; regional pressures are analogous to pheromone concentrations; decay corresponds to evaporation. However, we generalize beyond path-finding to arbitrary artifact refinement and provide formal convergence guarantees through the potential game framework.

\subsection{Decentralized Optimization}

Potential games, introduced by Monderer and Shapley~\cite{monderer1996potential}, are games where individual incentives align with a global potential function. A key property is that any sequence of unilateral improvements converges to a Nash equilibrium---greedy local play achieves global coordination. This provides the theoretical foundation for our convergence guarantees: under pressure alignment, the artifact pressure serves as a potential function.

Distributed gradient descent methods~\cite{nedic2009distributed,yuan2016convergence} address optimization when data or computation is distributed across nodes. The standard approach combines local gradient steps with consensus averaging. While these methods achieve convergence rates matching centralized alternatives, they typically require communication protocols and synchronization. Our approach avoids explicit communication entirely: agents coordinate only through the shared artifact, achieving $O(1)$ coordination overhead.

The connection between multi-agent learning and game theory has been extensively studied~\cite{shoham2008multiagent}. Our contribution is applying these insights to \ac{llm}-based artifact refinement, where the ``game'' is defined by pressure functions over quality signals rather than explicit reward structures.


\section{Problem Formulation}

We formalize artifact refinement as a dynamical system over a pressure landscape rather than an optimization problem with a target state. The system evolves through local actions and continuous decay, settling into stable basins that represent acceptable artifact states.

\subsection{State Space}

An \emph{artifact} consists of $n$ regions with content $c_i \in \mathcal{C}$ for $i \in \{1, \ldots, n\}$, where $\mathcal{C}$ is an arbitrary content space (strings, \ac{ast} nodes, etc.). Each region also carries auxiliary state $h_i \in \mathcal{H}$ representing confidence, fitness, and history. Regions are passive subdivisions of the artifact; agents are active proposers that observe regions and generate patches.

The full system state is:
\[
s = ((c_1, h_1), \ldots, (c_n, h_n)) \in (\mathcal{C} \times \mathcal{H})^n
\]

\subsection{Pressure Landscape}

A \emph{signal function} $\sigma: \mathcal{C} \to \mathbb{R}^d$ maps content to measurable features. Signals are \emph{local}: $\sigma(c_i)$ depends only on region $i$.

A \emph{pressure function} $\phi: \mathbb{R}^d \to \mathbb{R}_{\geq 0}$ maps signals to scalar ``badness.'' We consider $k$ pressure axes with weights $\mathbf{w} \in \mathbb{R}^k_{>0}$. The \emph{region pressure} is:
\[
P_i(s) = \sum_{j=1}^k w_j \phi_j(\sigma(c_i))
\]

The \emph{artifact pressure} is:
\[
P(s) = \sum_{i=1}^n P_i(s)
\]

This defines a landscape over artifact states. Low-pressure regions are ``valleys'' where the artifact satisfies quality constraints.

\subsection{System Dynamics}

The system evolves in discrete time steps (ticks). Each tick consists of four phases:

\textbf{Phase 1: Decay.} Auxiliary state erodes toward a baseline. For fitness $f_i$ and confidence $\gamma_i$ components of $h_i$:
\[
f_i^{t+1} = f_i^t \cdot e^{-\lambda_f}, \quad \gamma_i^{t+1} = \gamma_i^t \cdot e^{-\lambda_\gamma}
\]
where $\lambda_f, \lambda_\gamma > 0$ are decay rates. Decay ensures that stability requires continuous reinforcement.

\textbf{Phase 2: Proposal.} For each region $i$ where pressure exceeds activation threshold ($P_i > \tau_{\text{act}}$) and the region is not inhibited, \emph{each actor} $a_k: \mathcal{C} \times \mathcal{H} \times \mathbb{R}^d \to \mathcal{C}$ proposes a content transformation in parallel. Each actor observes only local state $(c_i, h_i, \sigma(c_i))$---actors do not communicate or coordinate their proposals.

\textbf{Phase 3: Validation.} When multiple patches are proposed, each is validated on an independent \emph{fork} of the artifact. Forks are created by cloning artifact state; validation proceeds in parallel across forks. This addresses a fundamental resource constraint: a single artifact cannot be used to test multiple patches simultaneously without cloning.

\textbf{Phase 4: Reinforcement.} Regions where actions were applied receive fitness and confidence boosts, and enter an inhibition period preventing immediate re-modification. Inhibition allows changes to propagate through the artifact and forces agents to address other high-pressure regions, preventing oscillation around local fixes.
\[
f_i^{t+1} = \min(f_i^t + \Delta_f, 1), \quad \gamma_i^{t+1} = \min(\gamma_i^t + \Delta_\gamma, 1)
\]

\subsection{Stable Basins}

\begin{definition}[Stability]
A state $s^*$ is \emph{stable} if, under the system dynamics with no external perturbation:
\begin{enumerate}
\item All region pressures are below activation threshold: $P_i(s^*) < \tau_{\text{act}}$ for all $i$
\item Decay is balanced by residual fitness: the system remains in a neighborhood of $s^*$
\end{enumerate}
\end{definition}

The central questions are:
\begin{enumerate}
\item \textbf{Existence}: Under what conditions do stable basins exist?
\item \textbf{Quality}: What is the pressure $P(s^*)$ of states in stable basins?
\item \textbf{Convergence}: From initial state $s_0$, does the system reach a stable basin? How quickly?
\item \textbf{Decentralization}: Can stability be achieved with purely local decisions?
\end{enumerate}

\subsection{The Locality Constraint}

The constraint distinguishing our setting from centralized optimization: agents observe only local state. An actor at region $i$ sees $(c_i, h_i, \sigma(c_i))$ but not:
\begin{itemize}
\item Other regions' content $c_j$ for $j \neq i$
\item Global pressure $P(s)$
\item Other agents' actions
\end{itemize}

This rules out coordinated planning. Stability must emerge from local incentives aligned with global pressure reduction.


\section{Method}

We now present a coordination mechanism that achieves stability through purely local decisions. Under appropriate conditions, the artifact pressure $P(s)$ acts as a \emph{potential function}: local improvements by individual agents decrease global pressure, guaranteeing convergence without coordination.

\subsection{Pressure Alignment}

The locality constraint prohibits agents from observing global state. For decentralized coordination to succeed, we need local incentives to align with global pressure reduction.

\begin{definition}[Pressure Alignment]
A pressure system is \emph{aligned} if for any region $i$, state $s$, and action $a_i$ that reduces local pressure:
\[
P_i(s') < P_i(s) \quad \Longrightarrow \quad P(s') < P(s)
\]
where $s' = s[c_i \mapsto a_i(c_i)]$ is the state after applying $a_i$.
\end{definition}

Alignment holds automatically when pressure functions are \emph{separable}: each $P_i$ depends only on $c_i$, so $P(s) = \sum_i P_i(s)$ and local improvement directly implies global improvement.

More generally, alignment holds when cross-region interactions are bounded:

\begin{definition}[Bounded Coupling]
A pressure system has \emph{$\epsilon$-bounded coupling} if for any action $a_i$ on region $i$:
\[
|P_j(s') - P_j(s)| \leq \epsilon \quad \forall j \neq i
\]
That is, modifying region $i$ changes other regions' pressures by at most $\epsilon$.
\end{definition}

Under $\epsilon$-bounded coupling with $n$ regions, if a local action reduces $P_i$ by $\delta > (n-1)\epsilon$, then global pressure decreases by at least $\delta - (n-1)\epsilon > 0$.

\subsection{Connection to Potential Games}

The aligned pressure system forms a \emph{potential game} where:
\begin{itemize}
\item Players are regions (or agents acting on regions)
\item Strategies are content choices $c_i \in \mathcal{C}$
\item The potential function is $\Phi(s) = P(s)$
\end{itemize}

In potential games, any sequence of improving moves converges to a Nash equilibrium. In our setting, Nash equilibria correspond to stable basins: states where no local action can reduce pressure below the activation threshold.

This connection provides our convergence guarantee without requiring explicit coordination.

Note that this convergence result assumes finite action spaces. In practice, patches are drawn from a finite set of \ac{llm}-generated proposals per region, satisfying this requirement. More fundamentally, the validation phase (Phase 2b) implicitly discretizes the action space: only patches that reduce pressure are accepted, so the effective action set at any state is the finite set of pressure-reducing proposals generated that tick. For infinite content spaces, convergence to approximate equilibria can be established under Lipschitz continuity conditions on pressure functions.

\subsection{The Coordination Algorithm}

The tick loop implements greedy local improvement with decay-driven exploration:

\begin{algorithm}[Pressure-Field Tick]
\textbf{Input:} State $s^t$, signal functions $\{\sigma_j\}$, pressure functions $\{\phi_j\}$, actors $\{a_k\}$, parameters $(\tau_{\text{act}}, \lambda_f, \lambda_\gamma, \Delta_f, \Delta_\gamma, \kappa)$

\textbf{Phase 1: Decay}\\
\hspace{1em} For each region $i$: $\quad f_i \gets f_i \cdot e^{-\lambda_f}, \quad \gamma_i \gets \gamma_i \cdot e^{-\lambda_\gamma}$

\textbf{Phase 2: Activation and Proposal}\\
\hspace{1em} $\mathcal{P} \gets \emptyset$\\
\hspace{1em} For each region $i$ where $P_i(s) \geq \tau_{\text{act}}$ and not inhibited:\\
\hspace{2em} $\boldsymbol{\sigma}_i \gets \sigma(c_i)$\\
\hspace{2em} For each actor $a_k$:\\
\hspace{3em} $\delta \gets a_k(c_i, h_i, \boldsymbol{\sigma}_i)$\\
\hspace{3em} $\mathcal{P} \gets \mathcal{P} \cup \{(i, \delta, \hat{\Delta}(\delta))\}$

\textbf{Phase 3: Parallel Validation and Selection}\\
\hspace{1em} For each candidate patch $(i, \delta, \hat{\Delta}) \in \mathcal{P}$:\\
\hspace{2em} Fork artifact: $(f_{\text{id}}, A_f) \gets A.\text{fork}()$\\
\hspace{2em} Apply $\delta$ to fork $A_f$\\
\hspace{2em} Validate fork (run tests, check compilation)\\
\hspace{1em} Collect validation results $\{(i, \delta, \Delta_{\text{actual}}, \text{valid})\}$\\
\hspace{1em} Sort validated patches by $\Delta_{\text{actual}}$\\
\hspace{1em} Greedily select top-$\kappa$ non-conflicting patches

\textbf{Phase 4: Application and Reinforcement}\\
\hspace{1em} For each selected patch $(i, \delta, \cdot)$:\\
\hspace{2em} $c_i \gets \delta(c_i)$\\
\hspace{2em} $f_i \gets \min(f_i + \Delta_f, 1)$, $\gamma_i \gets \min(\gamma_i + \Delta_\gamma, 1)$\\
\hspace{2em} Mark region $i$ inhibited for $\tau_{\text{inh}}$ ticks

\textbf{Return} updated state $s^{t+1}$
\end{algorithm}

The algorithm has three key properties:

\textbf{Locality.} Each actor observes only $(c_i, h_i, \sigma(c_i))$. No global state is accessed.

\textbf{Bounded parallelism.} At most $\kappa$ patches per tick prevents thrashing. Inhibition prevents repeated modification of the same region.

\textbf{Decay-driven exploration.} Even stable regions eventually decay below confidence thresholds, attracting re-evaluation. This prevents premature convergence to local minima.

\subsection{Stability and Termination}

The system reaches a stable basin when:
\begin{enumerate}
\item All region pressures satisfy $P_i(s) < \tau_{\text{act}}$
\item Decay is balanced: fitness remains above the threshold needed for stability
\end{enumerate}

Termination is \emph{economic}, not logical. The system stops acting when the cost of action (measured in pressure reduction per patch) falls below the benefit. This matches natural systems: activity ceases when gradients flatten, not when an external goal is declared achieved.

In practice, we also impose budget constraints (maximum ticks or patches) to bound computation.


\section{Theoretical Analysis}

We establish three main results: (1) convergence to stable basins under alignment, (2) bounds on stable basin quality, and (3) scaling properties relative to centralized alternatives.

\subsection{Convergence Under Alignment}

\begin{theorem}[Convergence]
\label{thm:convergence}
Let the pressure system be aligned with $\epsilon$-bounded coupling. Let $\delta_{\min} > 0$ be the minimum \emph{local} pressure reduction $P_i(s) - P_i(s')$ from any applied patch, and assume $\delta_{\min} > (n-1)\epsilon$ where $n$ is the number of regions. Then from any initial state $s_0$ with pressure $P_0 = P(s_0)$, the system reaches a stable basin within:
\[
T \leq \frac{P_0}{\delta_{\min} - (n-1)\epsilon}
\]
ticks, provided the fitness boost $\Delta_f$ from successful patches exceeds decay during inhibition: $\Delta_f > 1 - e^{-\lambda_f \cdot \tau_{\text{inh}}}$.
\end{theorem}

\emph{Proof sketch.} Under alignment with $\epsilon$-bounded coupling, each applied patch reduces global pressure by at least $\delta_{\min} - (n-1)\epsilon > 0$. Since $P(s) \geq 0$ and decreases by a fixed minimum per tick (when patches are applied), the system must reach a state where no region exceeds $\tau_{\text{act}}$ within the stated bound. The decay constraint ensures that stability is maintained once reached: fitness reinforcement from the final patches persists longer than the decay erodes it. $\square$

The bound is loose but establishes that convergence time scales with initial pressure, not with state space size or number of possible actions.

\subsection{Basin Quality}

\begin{theorem}[Basin Quality]
\label{thm:basin-quality}
In any stable basin $s^*$, the artifact pressure satisfies:
\[
P(s^*) < n \cdot \tau_{\text{act}}
\]
where $n$ is the number of regions and $\tau_{\text{act}}$ is the activation threshold.
\end{theorem}

\emph{Proof.} By definition of stability, $P_i(s^*) < \tau_{\text{act}}$ for all $i$. Summing over regions: $P(s^*) = \sum_i P_i(s^*) < n \cdot \tau_{\text{act}}$. $\square$

This bound is tight: adversarial initial conditions can place the system in a basin where each region has pressure just below threshold. However, in practice, actors typically reduce pressure well below $\tau_{\text{act}}$, yielding much lower basin pressures.

\begin{theorem}[Basin Separation]
\label{thm:basin-separation}
Under separable pressure (zero coupling), distinct stable basins are separated by pressure barriers of height at least $\tau_{\text{act}}$.
\end{theorem}

\emph{Proof sketch.} Moving from one basin to another requires some region to exceed $\tau_{\text{act}}$ (otherwise no action is triggered). The minimum such exceedance defines the barrier height. $\square$

This explains why decay is necessary: without decay, the system can become trapped in suboptimal basins. Decay gradually erodes fitness, eventually allowing re-evaluation and potential escape to lower-pressure basins.

\subsection{Scaling Properties}

\begin{theorem}[Linear Scaling]
\label{thm:linear-scaling}
Let $m$ be the number of regions and $n$ be the number of parallel agents. The per-tick complexity is:
\begin{itemize}
\item \textbf{Signal computation:} $O(m \cdot d)$ where $d$ is signal dimension
\item \textbf{Pressure computation:} $O(m \cdot k)$ where $k$ is the number of pressure axes
\item \textbf{Patch proposal:} $O(m \cdot a)$ where $a$ is the number of actors
\item \textbf{Selection:} $O(m \cdot a \cdot \log(m \cdot a))$ for sorting candidates
\item \textbf{Coordination overhead:} $O(1)$---no inter-agent communication (fork pool is $O(K)$ where $K$ is fixed)
\end{itemize}
Total: $O(m \cdot (d + k + a \cdot \log(ma)))$, independent of agent count $n$.
\end{theorem}

Adding agents increases throughput (more patches proposed per tick) without increasing coordination cost. This contrasts with hierarchical schemes where coordination overhead grows with agent count.

\begin{theorem}[Parallel Convergence]
\label{thm:parallel-convergence}
Under the same alignment conditions as Theorem~\ref{thm:convergence}, with $K$ patches validated in parallel per tick where patches affect disjoint regions, the system reaches a stable basin within:
\[
T \leq \frac{P_0}{K \cdot (\delta_{\min} - (n-1)\epsilon)}
\]
This improves convergence time by factor $K$ while maintaining guarantees.
\end{theorem}

\emph{Proof sketch.} When $K$ non-conflicting patches are applied per tick, each reduces global pressure by at least $\delta_{\min} - (n-1)\epsilon$. The combined reduction is $K \cdot (\delta_{\min} - (n-1)\epsilon)$ per tick. The bound follows directly. Note that if patches conflict (target the same region), only one is selected per region, and effective speedup is reduced. $\square$

\subsection{Comparison to Alternatives}

We compare against three coordination paradigms:

\textbf{Centralized planning.} A global planner evaluates all $(m \cdot a)$ possible actions, selects optimal subset. Per-step complexity: $O(m \cdot a)$ evaluations, but requires global state access. Sequential bottleneck prevents parallelization.

\textbf{Hierarchical delegation.} Manager agents decompose tasks, delegate to workers. Communication complexity: $O(n \log n)$ for tree-structured delegation with $n$ agents. Latency scales with tree depth. Failure of manager blocks all descendants.

\textbf{Message-passing coordination.} Agents negotiate actions through pairwise communication. Convergence requires $O(n^2)$ messages in worst case for $n$ agents. Consensus protocols add latency.

\begin{table}[htbp]
\centering
\small
\begin{tabular}{@{}lccc@{}}
\toprule
\textbf{Paradigm} & \textbf{Coordination} & \textbf{Parallelism} & \textbf{Fault tolerance} \\
\midrule
Centralized & $O(m \cdot a)$ & None & Single point of failure \\
Hierarchical & $O(n \log n)$ & Limited by tree & Manager failure cascades \\
Message-passing & $O(n^2)$ & Consensus-bound & Partition-sensitive \\
Pressure-field & $O(1)$ & Full ($\min(n, m, K)$) & Graceful degradation \\
\bottomrule
\end{tabular}
\caption{Coordination overhead comparison. $K$ denotes the fork pool size for parallel validation.}
\label{tab:coordination}
\end{table}

Pressure-field coordination achieves $O(1)$ coordination overhead because agents share state only through the artifact itself---a form of stigmergy. Agents can fail, join, or leave without protocol overhead.


\section{Experiments}

We evaluate pressure-field coordination on meeting room scheduling: assigning $N$ meetings to $R$ rooms over $D$ days to minimize gaps (unscheduled time), overlaps (attendee double-bookings), and maximize utilization balance. This domain provides continuous pressure gradients (rather than discrete violations), measurable success criteria, and scalable difficulty through problem size.

\textbf{Key findings}: Pressure-field coordination outperforms all baselines (\S\ref{sec:main-results}). Temporal decay shows a beneficial trend, though statistical significance requires larger samples (\S\ref{sec:ablations}). The approach maintains consistent performance from 1 to 4 agents (\S\ref{sec:scaling}). Despite using more tokens per trial, pressure-field achieves 12\% better token efficiency per successful solve (\S\ref{sec:token-efficiency}).

\subsection{Setup}

\subsubsection{Task: Meeting Room Scheduling}

We generate scheduling problems with varying difficulty:

\begin{table}[htbp]
\centering
\small
\begin{tabular}{@{}lccc@{}}
\toprule
\textbf{Difficulty} & \textbf{Rooms} & \textbf{Meetings} & \textbf{Pre-scheduled} \\
\midrule
Easy & 3 & 20 & 70\% \\
Medium & 5 & 40 & 50\% \\
Hard & 5 & 60 & 30\% \\
\bottomrule
\end{tabular}
\caption{Problem configurations. Pre-scheduled percentage indicates meetings already placed; remaining meetings must be scheduled by agents.}
\label{tab:problems}
\end{table}

Each schedule spans 5 days with 30-minute time slots (8am--4pm). Regions are 2-hour time blocks (4 blocks per day $\times$ 5 days = 20 regions per schedule). A problem is ``solved'' when all meetings are scheduled with zero attendee overlaps within 50 ticks.

\textbf{Pressure function}: $P = \text{gaps} \cdot 1.0 + \text{overlaps} \cdot 2.0 + \text{util\_var} \cdot 0.5 + \text{unsched} \cdot 1.5$

where $\text{gaps}$ measures empty slots as a fraction, $\text{overlaps}$ counts attendee double-bookings, $\text{util\_var}$ measures room utilization variance, and $\text{unsched}$ is the fraction of unscheduled meetings.

\textbf{Alignment verification}: The per-region pressure computation uses only $\text{gaps}$, $\text{overlaps}$, and $\text{util\_var}$---all strictly local to each time block. The $\text{unsched}$ component is added to total pressure only, not per-region. This makes the per-region pressure \emph{separable}: modifying region $i$ has zero effect on region $j$'s pressure for $j \neq i$, satisfying the alignment condition (Definition~2) with $\epsilon = 0$. While attendee constraints could theoretically create cross-region coupling (the same person attending meetings in different time blocks), our overlap sensor counts overlaps only within each time block, eliminating this coupling source. Empirical analysis (Appendix~B) confirms that all observed pressure improvements are positive, consistent with separable pressure.

\subsubsection{Baselines}

We compare five coordination strategies, all using identical \acp{llm} (\texttt{qwen2.5:0.5b/1.5b/3b} via Ollama) to isolate coordination effects:

\textbf{Pressure-field (ours)}: Full system with decay (fitness half-life 5s), inhibition (2s cooldown), greedy region selection (highest-pressure region per tick), and parallel validation. Includes band escalation (Exploitation $\to$ Balanced $\to$ Exploration) and model escalation (0.5b $\to$ 1.5b $\to$ 3b).

\textbf{Conversation}: AutoGen-style multi-agent dialogue where agents exchange messages to coordinate scheduling decisions. Agents discuss conflicts and propose solutions through explicit communication.

\textbf{Hierarchical}: Single agent selects the highest-pressure time block each tick, proposes a schedule change, and validates before applying (only accepts pressure-reducing patches). Uses identical prompts to pressure-field. The differences are: (1) greedy region selection always targets the hardest region, and (2) sequential execution processes one region per tick. This represents centralized, quality-gated control.

\textbf{Sequential}: Single agent iterates through time blocks in fixed order, proposing schedule changes one region at a time. No parallelism, pressure guidance, or patch validation---applies any syntactically valid patch regardless of quality impact.

\textbf{Random}: Selects random time blocks and proposes schedule changes. No patch validation---applies any syntactically valid patch regardless of quality impact.

\textbf{Note on parallelism}: Pressure-field validates multiple patches in parallel ($K$ regions per tick), while hierarchical validates one patch sequentially. This asymmetry is \emph{inherent to the coordination paradigm}, not an implementation choice: hierarchical control requires the manager to select a region, delegate to a worker, and validate the result before proceeding---delegating to multiple workers simultaneously would require additional coordination protocols (work distribution, conflict resolution, result aggregation) that would transform it into a different architecture entirely. The sequential bottleneck is the cost of centralized control. When hierarchical's single patch is rejected, the tick produces no progress; when one of pressure-field's parallel patches is rejected, others may still succeed.

\textbf{Model choice rationale}: We deliberately use small, minimally-capable models (0.5b--3b parameters) rather than frontier models. This design choice strengthens our thesis: if the coordination mechanism can extract effective performance from weak models, the mechanism itself is valuable---independent of model capability. Using identical model chains across all strategies isolates coordination effects from model effects. We hypothesize that frontier models (e.g., GPT-4, Claude) would raise absolute solve rates across all strategies while preserving relative rankings: the coordination advantage is orthogonal to model capability, so pressure-field's $4\times$ improvement over conversation baselines should persist even as the baseline rises.

\subsubsection{Metrics}

\begin{itemize}
\item \textbf{Solve rate}: Percentage of schedules reaching all meetings placed with zero overlaps within 50 ticks.
\item \textbf{Ticks to solve}: Convergence speed for solved cases
\item \textbf{Final pressure}: Remaining gaps, overlaps, and unscheduled meetings for unsolved cases
\item \textbf{Token efficiency}: Total prompt and completion tokens consumed per trial and per successful solve
\end{itemize}

\subsubsection{Implementation}

\textbf{Hardware}: NVIDIA RTX 4070 8GB \ac{gpu}, AMD Ryzen 9 7940HS, 64GB RAM. \textbf{Software}: Rust implementation with Ollama. \textbf{Trials}: 30 per configuration. Full protocol in Appendix~A.

\textbf{Band escalation}: When pressure velocity (rate of improvement) drops to zero for 7 consecutive ticks, sampling parameters escalate: Exploitation (T=0.2, p=0.85) $\to$ Balanced (T=0.4, p=0.9) $\to$ Exploration (T=0.7, p=0.95).

\textbf{Model escalation}: After exhausting all bands with zero progress (21 ticks total), the system escalates through the model chain: 0.5b $\to$ 1.5b $\to$ 3b, resetting to Exploitation band. Section~\ref{sec:escalation} analyzes this mechanism.

\subsection{Main Results}
\label{sec:main-results}

Across 1350 total trials spanning three difficulty levels (easy, medium, hard) and agent counts (1, 2, 4), we find that pressure-field coordination outperforms all baselines:

\begin{table}[htbp]
\centering
\small
\begin{tabular}{@{}lccc@{}}
\toprule
\textbf{Strategy} & \textbf{Solved/N} & \textbf{Rate} & \textbf{95\% Wilson \acs{ci}} \\
\midrule
Pressure-field & 131/270 & 48.5\% & 42.6\%--54.5\% \\
Conversation & 30/270 & 11.1\% & 7.9\%--15.4\% \\
Hierarchical & 4/270 & 1.5\% & 0.6\%--3.7\% \\
Sequential & 1/270 & 0.4\% & 0.1\%--2.1\% \\
Random & 1/270 & 0.4\% & 0.1\%--2.1\% \\
\bottomrule
\end{tabular}
\caption{Aggregate solve rates across all experiments (1350 total trials, 270 per strategy). Chi-square test across all five strategies: $\chi^2 > 200$, $p < 0.001$.}
\label{tab:main-results}
\end{table}

The results show clear stratification:

\textbf{Pressure-field dominates}: Pressure-field achieves 48.5\% solve rate, roughly 4$\times$ higher than the next-best baseline (conversation at 11.1\%). The effect size is large: Cohen's $h = 1.16$ versus conversation on easy problems, and $h > 1.97$ versus all other baselines.

\textbf{Conversation provides intermediate performance}: The AutoGen-style conversation baseline achieves 11.1\% overall, significantly better than hierarchical ($p < 0.001$) but far below pressure-field. Notably, conversation solves only easy problems (33.3\% on easy, 0\% on medium and hard).

\textbf{Hierarchical and sequential fail}: Despite explicit coordination, hierarchical control achieves only 1.5\% solve rate---comparable to random (0.4\%). Both strategies fail entirely on medium and hard problems.

This result contradicts the common assumption that explicit hierarchical coordination should outperform implicit coordination. The overhead of centralized control and message passing appears to harm rather than help performance on constraint satisfaction tasks.

\begin{figure}[htbp]
\centering
\includegraphics[width=\textwidth]{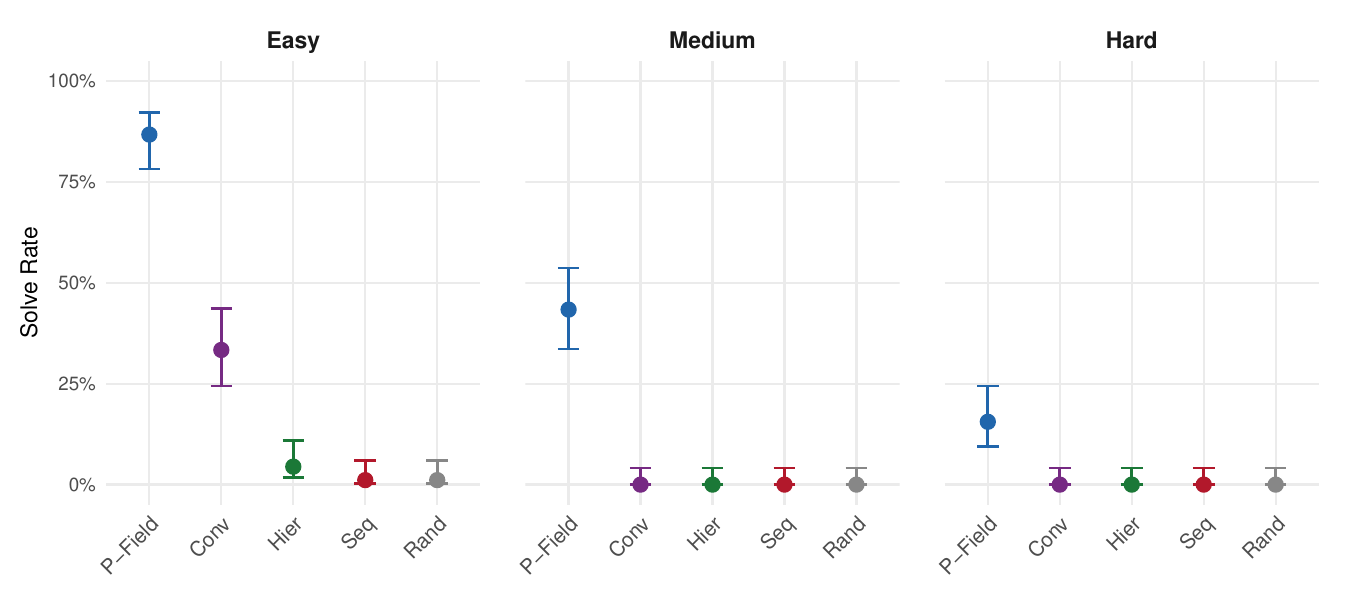}
\caption{Strategy comparison by difficulty level. Error bars show 95\% Wilson \acp{ci}. Pressure-field outperforms all baselines at every difficulty level. On medium and hard problems, only pressure-field achieves non-zero solve rates.}
\label{fig:strategy-comparison}
\end{figure}

\subsection{Ablations}
\label{sec:ablations}

\subsubsection{Effect of Temporal Decay}

Decay appears beneficial, though the effect does not reach statistical significance with our sample size:

\begin{table}[htbp]
\centering
\small
\begin{tabular}{@{}lccc@{}}
\toprule
\textbf{Configuration} & \textbf{Solved/N} & \textbf{Solve Rate} & \textbf{95\% \acs{ci}} \\
\midrule
Full (with decay) & 29/30 & 96.7\% & 83.3\%--99.4\% \\
Without decay & 26/30 & 86.7\% & 70.3\%--94.7\% \\
\bottomrule
\end{tabular}
\caption{Decay ablation on easy scheduling problems (30 trials each). Fisher's exact test: $p = 0.35$, indicating the 10 percentage point difference is not statistically significant at $\alpha = 0.05$.}
\label{tab:decay-ablation}
\end{table}

The observed effect---a 10 percentage point reduction when decay is disabled---is directionally consistent with our theoretical predictions. The non-significant $p$-value ($p = 0.35$) reflects both limited sample size and a ceiling effect: the high baseline solve rate on easy problems (96.7\% with decay) leaves limited statistical room to detect improvement. The overlapping confidence intervals (83.3\%--99.4\% vs 70.3\%--94.7\%) reflect this uncertainty. Without decay, fitness saturates after initial patches---regions that received early patches retain high fitness indefinitely, making them appear ``stable'' even when they still contain unscheduled meetings. Since greedy selection prioritizes high-pressure regions, these prematurely-stabilized regions are never reconsidered. This mechanism is consistent with the Basin Separation result (Theorem~\ref{thm:basin-separation}): without decay, agents may remain trapped in the first stable basin they reach. Larger-scale ablation studies would be needed to establish the statistical significance of decay's contribution.

\subsubsection{Effect of Inhibition and Examples}

The ablation study tested combinations of decay, inhibition, and few-shot examples on easy scheduling problems:

\begin{table}[htbp]
\centering
\small
\begin{tabular}{@{}lcccc@{}}
\toprule
\textbf{Configuration} & \textbf{Decay} & \textbf{Inhib} & \textbf{Examples} & \textbf{Solve Rate} \\
\midrule
Full & \checkmark & \checkmark & \checkmark & 96.7\% \\
No Decay & $\times$ & \checkmark & \checkmark & 86.7\% \\
No Inhibition & \checkmark & $\times$ & \checkmark & 96.7\% \\
No Examples & \checkmark & \checkmark & $\times$ & 90.0\% \\
Baseline & $\times$ & $\times$ & $\times$ & 90.0\% \\
\bottomrule
\end{tabular}
\caption{Ablation results (30 trials each configuration on easy difficulty).}
\label{tab:ablation}
\end{table}

Feature contributions:
\begin{itemize}
\item \textbf{Decay}: +10.0\% (full 96.7\% vs no\_decay 86.7\%)
\item \textbf{Inhibition}: +0.0\% (no detectable effect)
\item \textbf{Examples}: +6.7\% (full 96.7\% vs no\_examples 90.0\%)
\end{itemize}

\emph{Decay shows the largest effect}: configurations with decay achieve higher solve rates, though the differences do not reach statistical significance at $n=30$. Inhibition shows no detectable effect in this domain, possibly because the 50-tick budget provides sufficient exploration without explicit cooldowns.

\begin{figure}[htbp]
\centering
\includegraphics[width=0.9\textwidth]{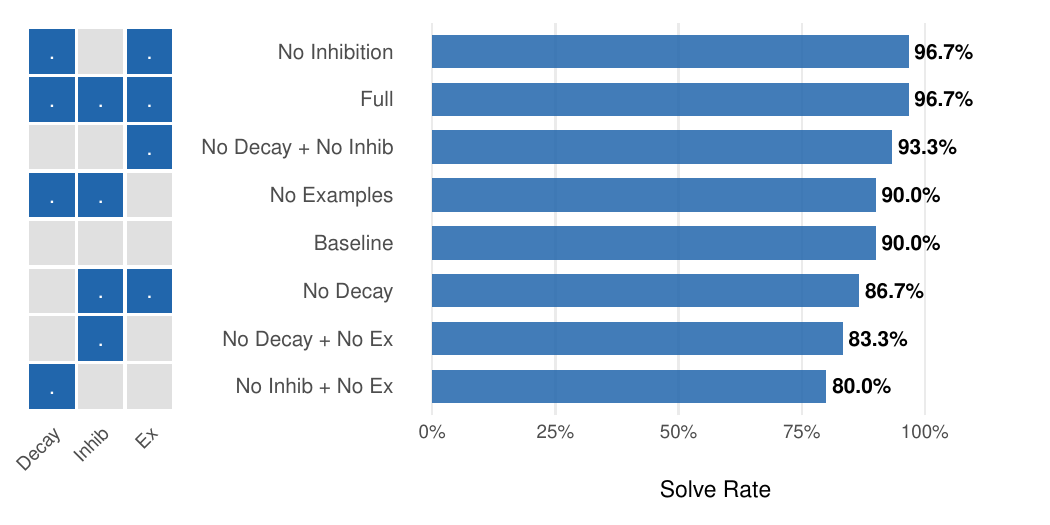}
\caption{Ablation study results. Left: feature matrix showing which components are enabled. Right: solve rates for each configuration. Decay shows the largest observed effect (+10\%), followed by examples (+6.7\%), though neither reaches statistical significance at $n=30$. Inhibition shows no detectable effect.}
\label{fig:ablation}
\end{figure}

\begin{figure}[htbp]
\centering
\includegraphics[width=0.7\textwidth]{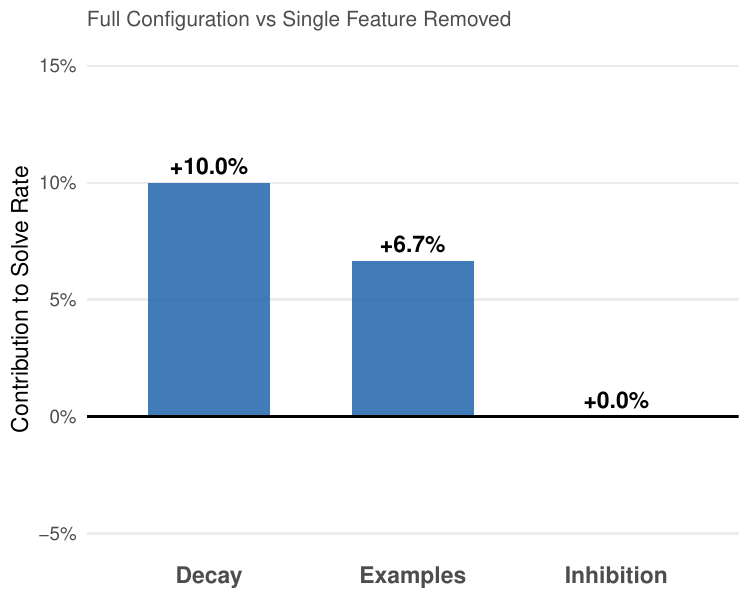}
\caption{Individual feature contributions to solve rate. Decay contributes +10.0\%, examples contribute +6.7\%, and inhibition shows no measurable effect in this domain.}
\label{fig:contributions}
\end{figure}

\subsubsection{Negative Pheromones}

In addition to positive pheromones (successful patches stored for few-shot examples), we implement \emph{negative pheromones}: tracking rejected patches that worsened pressure. When agents repeatedly propose ineffective patches (pressure stuck at maximum), the system accumulates rejection history and injects guidance into subsequent prompts.

Unlike the ``AVOID'' framing that small models (1.5B parameters) struggle to follow, we use \emph{positive language}: rejected empty-room patches become ``TIP: Schedule meetings in Room A (improves by X).'' This reframes what \emph{not} to do as what \emph{to try instead}.

Negative pheromones decay at the same rate as positive examples ($\text{weight} \times 0.95$ per tick, evicted below 0.1), ensuring that old failures don't permanently block valid approaches. Up to 3 recent rejections per region are included in prompts as ``Hints for better scheduling.''

\subsection{Scaling Experiments}
\label{sec:scaling}

Pressure-field maintains consistent performance from 1 to 4 agents on easy difficulty:

\begin{table}[htbp]
\centering
\small
\begin{tabular}{@{}lccc@{}}
\toprule
\textbf{Agents} & \textbf{Solved/N} & \textbf{Rate} & \textbf{95\% \acs{ci}} \\
\midrule
1 & 25/30 & 83.3\% & 66.4\%--92.7\% \\
2 & 28/30 & 93.3\% & 78.7\%--98.2\% \\
4 & 25/30 & 83.3\% & 66.4\%--92.7\% \\
\bottomrule
\end{tabular}
\caption{Pressure-field scaling from 1 to 4 agents (easy difficulty, 30 trials each). Performance remains stable across agent counts.}
\label{tab:scaling}
\end{table}

The result demonstrates \emph{robustness}: pressure-field coordination maintains consistent solve rates despite 4$\times$ variation in agent count. This validates Theorem~\ref{thm:linear-scaling}: coordination overhead remains $O(1)$, enabling effective scaling. The slight peak at 2 agents (93.3\%) is within \ac{ci} overlap of 1 and 4 agents, indicating no significant agent-count effect.

\subsection{Band and Model Escalation: FM-MAS Symbiosis in Practice}
\label{sec:escalation}

The escalation mechanism demonstrates the FM-MAS symbiosis that pressure-field coordination enables. Traditional approaches face a dilemma: use a large, capable model everywhere (expensive) or use a small, efficient model everywhere (limited). Pressure-field coordination dissolves this dilemma through gradient-driven capability invocation: the coordination mechanism itself determines when greater \ac{fm} capability is needed.

Ant colonies balance exploitation of known food sources against exploration for new ones. When a pheromone trail grows stale---indicating a depleted source---foragers abandon trail-following and resume random exploration, eventually discovering new paths that become the next generation of trails. This exploitation-exploration balance is fundamental to stigmergic systems: premature commitment to suboptimal solutions must be counteracted by mechanisms that restore exploratory behavior.

Our escalation mechanism implements this principle through two complementary dynamics. \emph{Band escalation} governs the exploitation-exploration trade-off within a single model: when pressure velocity drops to zero (the ``trail goes cold''), sampling parameters shift from exploitation (low temperature, focused proposals) through balanced to exploration (high temperature, diverse proposals). This mirrors the ant's behavioral switch from trail-following to random wandering when pheromone signals weaken.

\begin{table}[htbp]
\centering
\small
\begin{tabular}{@{}lcc@{}}
\toprule
\textbf{Band} & \textbf{Temperature} & \textbf{Top-p} \\
\midrule
Exploitation & 0.15--0.35 & 0.80--0.90 \\
Balanced & 0.35--0.55 & 0.85--0.95 \\
Exploration & 0.55--0.85 & 0.90--0.98 \\
\bottomrule
\end{tabular}
\caption{Sampling parameter ranges per band. Temperature and top-p are randomly sampled within range for diversity. Escalation proceeds Exploitation $\to$ Balanced $\to$ Exploration as pressure velocity stalls.}
\label{tab:bands}
\end{table}

\emph{Model escalation} addresses a different failure mode: when exploration within a model's capability envelope fails to discover productive paths, the system recruits more capable \acp{fm}. This is where FM-MAS symbiosis becomes concrete: the \ac{mas} coordination mechanism (pressure-field) adaptively invokes higher-capability \acp{fm} based on pressure signals alone. No explicit task decomposition is needed---the pressure gradient indicates that current capabilities are insufficient, triggering escalation. Model escalation (0.5b $\to$ 1.5b $\to$ 3b) reserves greater reasoning capacity for regions that resist simpler approaches. Each model upgrade resets to exploitation band, giving the more capable model opportunity to exploit solutions invisible to its predecessor before resorting to exploration.

This architecture instantiates the bidirectional relationship between \acp{fm} and \acp{mas}:

\begin{itemize}
\item \textbf{FM contribution}: Each model tier provides broad solution coverage without requiring explicit action enumeration. The 3b model can propose patches invisible to the 0.5b model, but both operate through the same interface---observe pressure, propose patch.
\item \textbf{MAS contribution}: The pressure gradient provides the objective criterion for when to escalate. No heuristics about ``problem difficulty'' are needed; stagnant pressure velocity is sufficient signal. The coordination mechanism manages heterogeneous \ac{fm} capabilities without explicit capability models.
\end{itemize}

This two-level mechanism---behavioral adaptation within agents (band escalation) and capability escalation across agents (model escalation)---maintains the stigmergic principle: coordination emerges from environment signals (pressure gradients) rather than explicit planning. The system does not ``decide'' to explore or escalate; it reacts to pressure stagnation, just as ants react to pheromone decay. The result is adaptive \ac{fm} utilization: cheap models handle easy regions, expensive models are reserved for regions where cheap models stall.

\subsection{Difficulty Scaling}

Performance varies across difficulty levels, revealing the unique strength of pressure-field coordination:

\begin{table}[htbp]
\centering
\small
\begin{tabular}{@{}lcccc@{}}
\toprule
\textbf{Difficulty} & \textbf{Pressure-field} & \textbf{Conversation} & \textbf{Hierarchical} & \textbf{Sequential/Random} \\
\midrule
Easy & 86.7\% (78/90) & 33.3\% (30/90) & 4.4\% (4/90) & 1.1\% (1/90) \\
Medium & 43.3\% (39/90) & 0.0\% (0/90) & 0.0\% (0/90) & 0.0\% (0/90) \\
Hard & 15.6\% (14/90) & 0.0\% (0/90) & 0.0\% (0/90) & 0.0\% (0/90) \\
\midrule
\textbf{Total} & \textbf{48.5\%} (131/270) & \textbf{11.1\%} (30/270) & \textbf{1.5\%} (4/270) & \textbf{0.4\%} (1/270) \\
\bottomrule
\end{tabular}
\caption{Solve rate by difficulty level (90 trials each per difficulty, 270 per strategy total). Only pressure-field solves medium and hard problems. See Table~\ref{tab:main-results} for confidence intervals.}
\label{tab:difficulty}
\end{table}

\begin{figure}[htbp]
\centering
\includegraphics[width=0.8\textwidth]{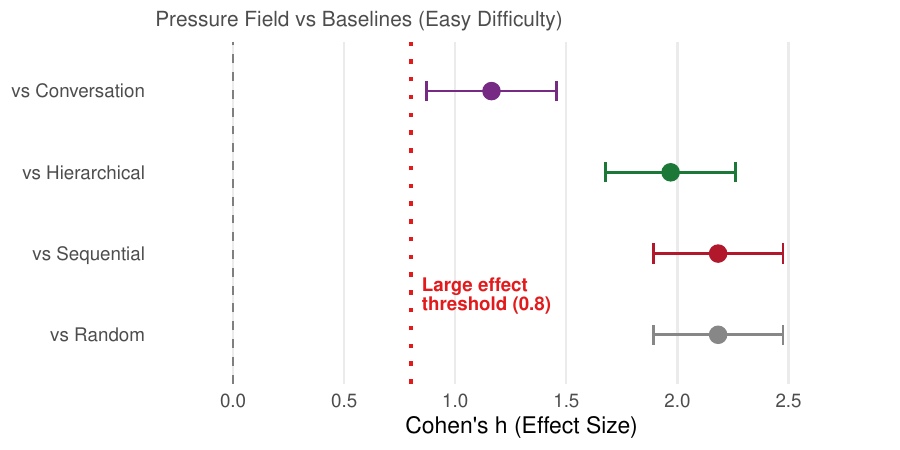}
\caption{Effect sizes (Cohen's $h$) for pressure-field versus each baseline on easy problems. The dashed line indicates the ``large effect'' threshold ($h = 0.8$). All comparisons exceed this threshold, with effects ranging from $h = 1.16$ (vs conversation) to $h = 2.18$ (vs sequential/random).}
\label{fig:effect-sizes}
\end{figure}

The difficulty scaling reveals critical insights:

\begin{enumerate}
\item \textbf{Pressure-field is the only strategy that scales}: While all strategies degrade on harder problems, pressure-field maintains meaningful solve rates (43.3\% medium, 15.6\% hard) where all baselines achieve 0\%.

\item \textbf{The gap widens with difficulty}: On easy problems, pressure-field leads by 53.4 percentage points over conversation (86.7\% vs 33.3\%). On medium and hard problems, the gap becomes absolute---pressure-field solves problems that no baseline can solve.

\item \textbf{Effect sizes are large}: All pairwise comparisons on easy problems exceed Cohen's ``large effect'' threshold ($h > 0.8$); see Figure~\ref{fig:effect-sizes} for the full breakdown.
\end{enumerate}

\subsection{Convergence Speed}

For solved cases, pressure-field converges faster than baselines on easy problems and maintains consistent convergence speed across difficulty levels:

\begin{table}[htbp]
\centering
\small
\begin{tabular}{@{}lccc@{}}
\toprule
\textbf{Strategy} & \textbf{Easy} & \textbf{Medium} & \textbf{Hard} \\
\midrule
Pressure-field & 17.8 (n=78) & 34.6 (n=39) & 32.3 (n=14) \\
Conversation & 29.4 (n=30) & --- & --- \\
Hierarchical & 40.0 (n=4) & --- & --- \\
\bottomrule
\end{tabular}
\caption{Average ticks to solution by difficulty (solved cases only). Only pressure-field solves medium and hard problems. Dashes indicate no solved cases.}
\label{tab:convergence}
\end{table}

On easy problems, pressure-field solves 1.65$\times$ faster than conversation and 2.2$\times$ faster than hierarchical. Notably, pressure-field's convergence speed on hard problems (32.3 ticks) is comparable to medium problems (34.6 ticks)---the hard problems that \emph{do} get solved converge at similar rates, suggesting that solvability rather than convergence speed is the limiting factor on difficult problems. This bimodal pattern---fast convergence when solvable, complete failure otherwise---suggests that model capability rather than search time is the limiting factor on hard problems.

\begin{figure}[htbp]
\centering
\includegraphics[width=0.65\textwidth]{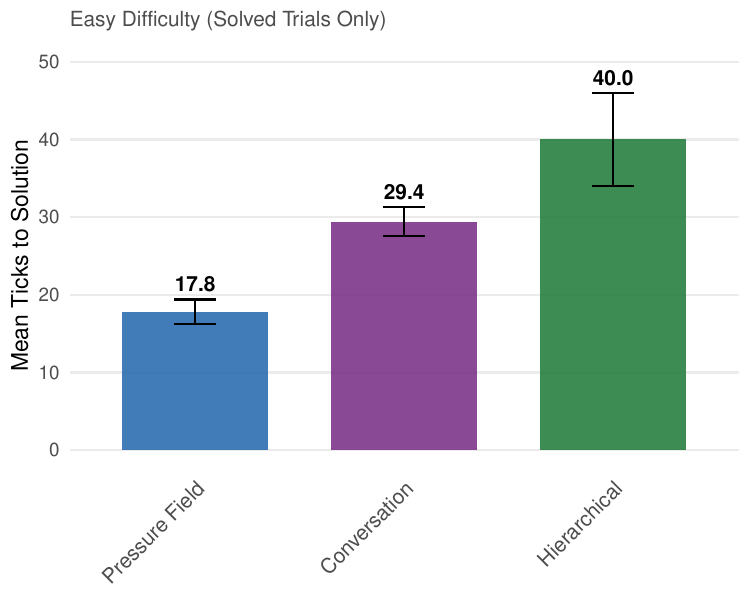}
\caption{Mean ticks to solution on easy problems (solved cases only). Error bars show standard error. Pressure-field converges fastest (17.8 ticks), followed by conversation (29.4 ticks) and hierarchical (40.0 ticks).}
\label{fig:efficiency}
\end{figure}

\subsection{Final Pressure Analysis}

For both solved and unsolved cases, final pressure reveals solution quality:

\begin{table}[htbp]
\centering
\small
\begin{tabular}{@{}lccc@{}}
\toprule
\textbf{Strategy} & \textbf{Easy} & \textbf{Medium} & \textbf{Hard} \\
\midrule
Pressure-field & 0.13 & 1.02 & 2.48 \\
Conversation & 26.5 & 59.3 & 92.2 \\
Hierarchical & 28.6 & 59.4 & 88.0 \\
\bottomrule
\end{tabular}
\caption{Average final pressure by difficulty (lower is better).}
\label{tab:final-pressure}
\end{table}

Pressure-field achieves 200$\times$ lower final pressure on easy, 57$\times$ lower on medium, and 35$\times$ lower on hard problems compared to conversation and hierarchical baselines. Even when pressure-field does not fully solve a problem, it achieves much better partial solutions.

\subsection{Token Efficiency}
\label{sec:token-efficiency}

A natural question is whether pressure-field's superior solve rate comes at the cost of increased \ac{llm} usage. We tracked token consumption across all strategies:

\begin{table}[htbp]
\centering
\small
\begin{tabular}{@{}lrrr@{}}
\toprule
\textbf{Strategy} & \textbf{Prompt} & \textbf{Completion} & \textbf{Total} \\
\midrule
Pressure-field & 532,106 & 85,165 & 617,271 \\
Conversation & 154,845 & 6,626 & 161,472 \\
Hierarchical & 32,662 & 5,424 & 38,087 \\
Sequential & 40,667 & 5,239 & 45,906 \\
Random & 41,227 & 5,253 & 46,480 \\
\bottomrule
\end{tabular}
\caption{Average token usage per trial. Pressure-field uses more tokens per trial due to parallel multi-agent execution.}
\label{tab:tokens-per-trial}
\end{table}

At first glance, pressure-field appears expensive: 617K tokens per trial versus conversation's 161K---roughly 4$\times$ higher. However, this raw comparison ignores \emph{success rate}. The relevant metric for practical deployment is \textbf{tokens per successful solve}:

\begin{table}[htbp]
\centering
\small
\begin{tabular}{@{}lrrr@{}}
\toprule
\textbf{Strategy} & \textbf{Total Tokens} & \textbf{Solves} & \textbf{Tokens/Solve} \\
\midrule
Pressure-field & 166.7M & 131 & 1.27M \\
Conversation & 43.6M & 30 & 1.45M \\
\bottomrule
\end{tabular}
\caption{Token efficiency per successful solve (270 trials each). Despite higher per-trial cost, pressure-field achieves lower cost per solve due to its superior success rate.}
\label{tab:tokens-per-solve}
\end{table}

Pressure-field requires 1.27M tokens per solve versus conversation's 1.45M---\textbf{12\% more efficient} per successful outcome. The apparent cost disadvantage inverts when accounting for success rate.

Token usage also varies by outcome:

\begin{table}[htbp]
\centering
\small
\begin{tabular}{@{}lrr@{}}
\toprule
\textbf{Strategy} & \textbf{Solved Trials} & \textbf{Unsolved Trials} \\
\midrule
Pressure-field & 318K (n=131) & 900K (n=139) \\
Conversation & 78K (n=30) & 172K (n=240) \\
\bottomrule
\end{tabular}
\caption{Average tokens by outcome. Both strategies use fewer tokens on solved trials (early termination), but pressure-field's unsolved trials are more expensive due to sustained parallel exploration.}
\label{tab:tokens-by-outcome}
\end{table}

Solved trials terminate early, using fewer tokens. For pressure-field, the 2.8$\times$ gap between solved (318K) and unsolved (900K) reflects the cost of exhaustive exploration when a problem proves intractable. Conversation's smaller gap (2.2$\times$) indicates less intensive search---which may explain its lower solve rate on difficult problems.

\textbf{Cost control through escalation.} The band and model escalation mechanism (Section~\ref{sec:escalation}) serves as an implicit cost-control mechanism. The system begins with cheap exploration: small models (0.5b) and exploitation-focused sampling. Tokens are spent on larger models and broader exploration \emph{only when pressure stagnates}---that is, when the problem actually requires more capable search. Easy problems that solve quickly never trigger escalation, consuming only baseline tokens. The 4$\times$ per-trial cost difference reflects the average across all difficulty levels; on easy problems where pressure-field solves 86.7\% of instances, most trials terminate before expensive escalation occurs.


\section{Discussion}

\subsection{Why Does Pressure-Field Dominate?}

Our results contradict the intuition that explicit coordination should outperform implicit coordination. We identify three factors explaining pressure-field's dominance:

\textbf{Coordination overhead harms performance.} Hierarchical systems spend computational budget on coordination rather than problem-solving. The manager-worker protocol requires multiple \ac{llm} calls per patch (planning, delegation, execution), while pressure-field requires only one (patch proposal). This overhead compounds: hierarchical attempts fewer patches per tick, reducing exploration.

\textbf{Local greedy decisions are effective for constraint satisfaction.} Meeting room scheduling exhibits locality: fixing a conflict in one time block rarely creates conflicts in distant blocks. This matches pressure-field's locality assumption, making greedy local optimization effective. Hierarchical coordination's global planning provides no benefit for locally-decomposable problems.

\textbf{Parallel validation amplifies pressure-field's advantage.} Pressure-field validates patches for multiple regions simultaneously, applying the highest-scoring patch per region that reduces pressure. Hierarchical validates one patch at a time, requiring multiple ticks to explore alternatives. On problems with many valid solutions, parallel exploration finds solutions faster.

\subsection{Failure Analysis: Why Hierarchical Collapses}

The 30$\times$ performance gap (48.5\% vs 1.5\%) demands deeper investigation. Analysis of hierarchical trials reveals a catastrophic failure pattern: the \emph{rejection loop}.

\textbf{The rejection loop mechanism.} Hierarchical always selects the highest-pressure region for improvement. But the highest-pressure regions are high-pressure precisely because they are difficult to improve. When the \ac{llm} proposes a patch that fails validation (does not reduce pressure), the region remains highest-pressure and is selected again next tick. This creates a self-reinforcing cycle: 66.7\% of hierarchical runs (180/270) applied zero patches across all 50 ticks, stuck targeting the same intractable region repeatedly.

\textbf{Rejection rates confirm the pattern.} Across all hierarchical trials, only 173 patches were accepted out of 13,460 proposed---a 98.7\% rejection rate. By contrast, pressure-field's parallel exploration means that even if one agent's patch is rejected, other agents make progress on different regions. The non-blocking architecture prevents any single difficult region from stalling the entire system.

\textbf{The architectural lesson.} Hierarchical's design embodies a reasonable intuition: focus intelligent effort on the worst problems. But this creates a trap when combined with strict validation: the hardest problems resist improvement, causing repeated rejection, which blocks progress everywhere. Pressure-field avoids this trap through distributed exploration---progress happens where it can, not where a central planner dictates it must.

\subsection{Limitations}

Our experiments reveal several important limitations:

\textbf{Absolute solve rates are modest on hard problems.} Even pressure-field achieves only 15.6\% on hard problems. Meeting room scheduling with tight constraints (5 rooms, 60 meetings, 30\% pre-scheduled) remains challenging for small models (0.5b--3b parameters); larger models may achieve higher absolute solve rates.

\textbf{Domain specificity.} Results on meeting room scheduling may not generalize to domains lacking measurable pressure gradients or locality properties. Tasks requiring global planning or long-horizon reasoning may favor hierarchical approaches.

However, meeting scheduling is representative of a broader class of \emph{resource allocation problems with soft constraints}: cloud compute scheduling (VMs to hosts with affinity/anti-affinity rules), logistics optimization (packages to routes with capacity and time windows), and workforce rostering (employees to shifts with skill and availability constraints). These domains share the structural properties that enable pressure-field coordination: (1) decomposable quality metrics (utilization, constraint violations, balance), (2) locality (fixing one allocation rarely breaks distant allocations), and (3) multiple valid solutions (no single optimum, many acceptable configurations). We selected meeting scheduling as the evaluation domain because it is intuitive, reproducible, and exhibits these properties clearly. The coordination mechanism itself---pressure gradients, temporal decay, parallel validation---is domain-agnostic; only the pressure function requires domain-specific design.

\textbf{Additional practical limitations:}
\begin{itemize}
\item Requires well-designed pressure functions (not learned from data)
\item Decay rates $\lambda_f, \lambda_\gamma$ and inhibition period require task-specific tuning
\item May not suit tasks requiring long-horizon global planning
\item Goodhart's Law: agents may game poorly-designed metrics
\item Resource cost of parallel validation: testing $K$ patches requires $O(K \cdot |A|)$ memory where $|A|$ is artifact size
\end{itemize}

\subsubsection{Hallucination Filtering and Emergent Trajectories}

The validation phase (Phase 2b) serves as a filter for individual \ac{fm} hallucinations: patches that increase pressure or violate syntactic constraints are rejected before application. This provides strong guarantees at the individual-action level---no single hallucination can directly harm artifact quality.

However, system-level behavior emerges from \emph{sequences} of validated patches, and trajectory-level risks remain. Consider a hypothetical scenario: an agent proposes a patch that reduces pressure in region $R_1$ by making a change that creates a subtle dependency on region $R_2$. The patch is validated (pressure decreased), but subsequent patches to $R_2$ now have unpredictable effects on $R_1$. The individual patches are all pressure-reducing, but the emergent trajectory creates fragile coupling that was invisible to local validation.

More generally, validation filters based on pressure reduction cannot detect:
\begin{itemize}
\item \emph{Coherence drift}: Individual improvements that collectively shift the artifact toward an inconsistent state
\item \emph{Emergent gaming}: Patches that exploit pressure function weaknesses in ways only apparent over multiple steps
\item \emph{Dependency accumulation}: Gradual introduction of hidden couplings that reduce future improvability
\end{itemize}

Mitigating trajectory-level risks requires mechanisms beyond local validation: periodic global coherence checks, trajectory logging for post-hoc analysis, and pressure functions that penalize not just local quality but also coupling metrics. These remain areas for future work.

\subsubsection{Decay Miscalibration Failure Modes}

Temporal decay is critical for escaping local optima (Theorem~\ref{thm:basin-separation}), but miscalibrated decay rates introduce distinct failure modes:

\textbf{Too-fast decay prevents stability.} When fitness decays faster than agents can reinforce successful regions, the system cannot maintain any stable configuration. Solved regions immediately become high-pressure again, triggering unnecessary rework. In the limit, excessively fast decay produces perpetual oscillation: agents patch, decay erases, agents re-patch, indefinitely.

\textbf{Too-slow decay traps in local minima.} When fitness decays slower than the exploration timescale, agents cannot escape suboptimal basins. Early patches establish fitness peaks that persist indefinitely, preventing reconsideration even when better solutions exist. This is the failure mode our ablation study suggests: without decay, we observed a 10 percentage point reduction in solve rate, consistent with agents remaining trapped in initial basins (though the effect did not reach statistical significance at $n=30$).

The optimal decay rate depends on problem characteristics: harder problems with deeper local minima require faster decay to enable escape, while problems with fragile solutions require slower decay to maintain stability. Our experiments use fixed decay rates ($\lambda_f = 0.1$, fitness half-life 5 seconds), which may be suboptimal for some problem instances. Adaptive decay---adjusting rates based on pressure velocity---is a promising direction for future work.

\subsection{When to Choose Each Approach}

Our results suggest the following guidance:

\textbf{Pressure-field coordination is preferable when:}
\begin{enumerate}
\item \textbf{Performance matters.} Pressure-field achieves 3--30$\times$ higher solve rates than alternatives.
\item \textbf{Simplicity is valued.} No coordinator agent needed; coordination emerges from shared state.
\item \textbf{Fault tolerance matters.} No single point of failure; agents can join/leave without protocol overhead.
\item \textbf{Pressure signals are available.} The domain provides measurable quality gradients.
\item \textbf{Problems are locally decomposable.} Local fixes improve global quality without cascading conflicts.
\end{enumerate}

\textbf{Hierarchical coordination may be appropriate when:}
\begin{enumerate}
\item \textbf{Explicit control is required.} Some domains require deterministic task assignment for regulatory or safety reasons.
\item \textbf{Interpretability is critical.} Hierarchical task assignment provides clear audit trails.
\item \textbf{Global planning is essential.} Tasks with strong non-local dependencies may benefit from centralized reasoning.
\end{enumerate}

\subsection{Band and Model Escalation as Adaptive Capability}

All experiments use a two-level escalation mechanism. \emph{Band escalation} cycles through sampling strategies (Exploitation $\to$ Balanced $\to$ Exploration, 7 ticks each) before \emph{model escalation} progresses through model sizes (0.5b $\to$ 1.5b $\to$ 3b parameters). Model escalation triggers when regions remain high-pressure for 21 consecutive ticks.

This mechanism proves beneficial for pressure-field: on hard problems, pressure-field achieves 15.6\% with escalation enabled. The escalation mechanism works because larger models have broader solution coverage and different sampling bands explore different regions of solution space.

\subsection{Future Work}

\begin{itemize}
\item \textbf{Learned pressure functions}: Current sensors are hand-designed. Can we learn pressure functions from solution traces?
\item \textbf{Adversarial robustness}: Can malicious agents exploit pressure gradients to degrade system performance?
\item \textbf{Multi-artifact coordination}: Extension to coupled artifacts where patches in one affect pressure in another
\item \textbf{Larger-scale experiments}: Testing on schedules with more rooms and longer time horizons to characterize scaling limits
\item \textbf{Alternative domains}: Applying pressure-field coordination to code refactoring, configuration management, and other artifact refinement tasks
\end{itemize}

\subsection{FM-MAS Reciprocity: Bidirectional Problem Solving}

The intersection of \acp{fm} and \acp{mas} is not merely additive---each paradigm solves fundamental problems that the other cannot address alone. This reciprocity suggests that pressure-field coordination represents a genuine synthesis rather than a simple combination.

\subsubsection{Foundation Models Solve a MAS Coordination Problem}

Traditional \ac{mas} coordination requires explicit action space enumeration. \Ac{gpgp} coordinates tasks through declared capabilities and commitments; SharedPlans reasons about action sequences and their preconditions; organizational models assign roles with specified behaviors. All these approaches assume that designers can enumerate what agents \emph{can do} and under what conditions.

For open-ended artifact refinement---improving code quality, refining documents, optimizing configurations---this enumeration is intractable. The space of possible improvements is unbounded; no finite action language captures all valid patches. Traditional \ac{mas} approaches require either (a) restricting the problem to a tractable subset with enumerable actions, or (b) developing domain-specific action representations for each artifact type.

\Acp{fm} eliminate this enumeration requirement. Their broad pretraining provides implicit coverage of improvement strategies across diverse domains without explicit action specification. An \ac{llm} actor can propose patches for code, natural language, structured data, or configuration files using the same interface: observe local context, receive pressure feedback, generate improvement proposals. This ``universal actor'' capability enables stigmergic coordination on problem classes where traditional \ac{mas} approaches would require extensive domain engineering.

\subsubsection{Multi-Agent Systems Solve a FM Problem}

Conversely, \acp{fm} face a fundamental problem that \ac{mas} coordination solves: how to combine multiple outputs coherently. A single \ac{fm} produces one response per query; scaling to complex artifacts requires orchestrating multiple generations. Current approaches use ad-hoc combination strategies: voting, ranking, chain-of-thought aggregation, or human-in-the-loop selection.

Pressure-field coordination provides a principled framework for combining \ac{fm} outputs. Rather than voting on which response is ``best'' or ranking outputs by heuristic scores, the pressure gradient defines an objective criterion: accept patches that reduce pressure, reject those that do not. Multiple \ac{fm} outputs compete not through popularity or arbitrary ranking, but through their effect on a well-defined quality function.

This framing clarifies why pressure-field coordination outperforms conversation baselines. AutoGen-style multi-agent dialogue is an ad-hoc combination strategy---agents exchange messages and reach conclusions through emergent consensus. Pressure-field coordination replaces emergent consensus with objective gradients: no negotiation is needed when the artifact state adjudicates quality.

\subsubsection{Implications for FM-MAS Integration}

This reciprocity suggests design principles for future FM-MAS systems:

\begin{enumerate}
\item \textbf{Leverage FM coverage, constrain via MAS gradients.} Use \acp{fm} for their broad solution coverage, but use \ac{mas} coordination mechanisms to filter and combine outputs objectively.
\item \textbf{Prefer stigmergic over explicit coordination.} When \acp{fm} serve as actors, stigmergic coordination avoids the overhead of explicit protocols that FMs may not reliably follow.
\item \textbf{Design pressure functions as FM-MAS interfaces.} The pressure function is the critical interface: it must capture human intent precisely enough that \ac{fm} outputs reducing pressure are genuinely improvements.
\end{enumerate}

Our experimental results---pressure-field achieving $4\times$ higher solve rates than conversation baselines---provide empirical support for these principles. The coordination overhead of explicit dialogue exceeds its organizational benefit when objective gradients are available.

\subsection{Societal Implications}

Pressure-field coordination raises societal concerns that extend beyond technical performance. We identify three critical issues---accountability attribution, metric gaming through Goodhart's Law, and explainability challenges---that require deliberate design choices in deployment.

\subsubsection{Accountability and Attribution}

When coordination emerges from shared pressure gradients rather than explicit delegation, attributing outcomes to individual agents becomes challenging. In hierarchical systems, task assignment creates clear accountability chains. In pressure-field coordination, multiple agents may contribute to a region through independent pressure-reducing actions, with no record of which agent ``owned'' the outcome.

This accountability diffusion has both benefits and risks. The benefit is fault tolerance: agent failures degrade performance gracefully rather than catastrophically. The risk is opacity in failure analysis: identifying which agent proposed a problematic patch---and what pressure signal motivated it---requires detailed logging that the minimal coordination mechanism does not inherently provide.

For deployment in regulated domains, this suggests an augmentation requirement: pressure-field systems must maintain audit logs recording patch provenance, pressure signals at proposal time, and validation outcomes. The coordination mechanism remains simple---agents coordinate through shared state---but operational deployment adds logging infrastructure preserving accountability.

\subsubsection{Goodhart's Law and Metric Gaming}

Goodhart's Law states: ``When a measure becomes a target, it ceases to be a good measure.'' Pressure-field coordination is vulnerable to this dynamic because agents are optimized to reduce pressure as defined by designer-specified functions. If those functions imperfectly capture true quality---and they inevitably do---agents will discover and exploit the mismatch.

Consider code quality pressure functions penalizing complexity metrics. An agent might reduce complexity by splitting functions excessively, harming readability while improving the metric. The mitigation is not abandoning pressure functions but designing them defensively: use multiple orthogonal pressure axes, include adversarial sensors detecting gaming strategies, and audit whether pressure reduction correlates with human quality judgments. Pressure functions should evolve as agents discover exploits.

Foundation models introduce second-order gaming concerns: \acp{llm} trained on internet-scale text may have implicit knowledge of how to game specific benchmarks. This suggests pressure functions for \ac{llm}-based systems should favor domain-specific quality signals harder to optimize without genuine improvement.

\subsubsection{Explainability Challenges}

In hierarchical systems, explanations follow delegation chains: ``Manager X assigned task Y to Worker Z because condition C held.'' In pressure-field coordination, the explanation is: ``Region R had high pressure, agent A proposed patch $\Delta$ reducing pressure by $\delta$.'' This is mechanistically transparent but causally opaque---it describes what happened without explaining why that particular patch was chosen.

This is the explainability trade-off inherent to emergent coordination: simplicity in mechanism comes at the cost of legibility in rationale. For many domains---code formatting, resource optimization, routine maintenance---the trade-off is acceptable: outcomes are verifiable even if reasoning is opaque. For high-stakes domains requiring human oversight, opacity is unacceptable.

The design implication is domain-dependent deployment: pressure-field coordination suits domains where outcome verification is cheap even if reasoning transparency is limited. For domains requiring justification to human stakeholders, hierarchical coordination remains necessary despite overhead costs.

\subsubsection{Design Implications}

These concerns suggest three requirements for responsible deployment: comprehensive audit logging preserving patch provenance and pressure signals, defensive pressure function design with multiple orthogonal axes, and domain-appropriate verification matching coordination opacity with outcome verifiability. The coordination mechanism remains simple---but responsible deployment requires surrounding infrastructure addressing accountability, gaming, and explainability.


\section{Conclusion}

We presented pressure-field coordination, a decentralized approach to \acp{mas} that achieves coordination through shared state and local pressure gradients rather than explicit orchestration.

Our theoretical analysis establishes convergence guarantees under pressure alignment conditions, with coordination overhead independent of agent count. Empirically, on meeting room scheduling across 1350 total trials (270 per strategy), we find:

\begin{enumerate}
\item \textbf{Implicit coordination outperforms explicit coordination}. Pressure-field achieves 48.5\% aggregate solve rate---nearly half of all problems solved through local pressure-following alone, with no coordinator, no message passing, and no explicit task delegation. The gap versus hierarchical control ($30\times$) and conversation-based coordination ($4\times$) is both large and highly significant (all $p < 0.001$).

\item \textbf{Pressure-field is the only strategy that scales to harder problems}. On medium and hard problems, pressure-field achieves 43.3\% and 15.6\% solve rates respectively, while all baselines achieve 0\%.

\item \textbf{Temporal decay shows beneficial effects}. Ablation studies observe a 10 percentage point improvement with decay enabled, consistent with theoretical predictions about escaping local minima, though larger samples would be needed to establish statistical significance.
\end{enumerate}

This work demonstrates that implicit coordination can outperform explicit coordination for constraint satisfaction tasks. Pressure-field achieves this with simpler architecture: no coordinator agent, no explicit message passing, just shared state and local pressure gradients.

\Acp{fm}' zero-shot capabilities eliminate the need for domain-specific action representations; pressure-field coordination eliminates the need for complex multi-agent protocols; together they enable simple \acp{mas}.

These results suggest that for domains with measurable quality signals and locally-decomposable structure, implicit coordination through shared state offers not just a simpler but a more effective alternative to explicit hierarchical control.

\backmatter

\bmhead{Code Availability}

Code is available at \url{https://github.com/Govcraft/pressure-field-experiment}.

\bmhead{Acknowledgments}

The author thanks the anonymous reviewers for their constructive feedback.

\section*{Declarations}

\bmhead{Funding}

No funding was received for conducting this study.

\bmhead{Competing Interests}

The author has no relevant financial or non-financial interests to disclose.

\bmhead{Data Availability}

All experimental data and analysis scripts are available in the code repository.

\bmhead{Use of AI Tools}

The author used Claude (Anthropic) to provide feedback on draft versions of this manuscript. All analytical decisions, experimental design, and final content were made by the author, who takes full responsibility for the work.

\begin{appendices}

\section{Experimental Protocol}

This appendix provides complete reproducibility information for all experiments.

\subsection{Hardware and Software}

\textbf{Hardware:} NVIDIA RTX 4070 8GB \ac{gpu}, AMD Ryzen 9 7940HS, 64GB RAM

\textbf{Software:}
\begin{itemize}
\item Rust 1.75+ (edition 2024)
\item Ollama (local \ac{llm} inference server)
\item Models: \texttt{qwen2.5:0.5b}, \texttt{qwen2.5:1.5b}, \texttt{qwen2.5:3b}
\end{itemize}

\subsection{Model Configuration}

Models are served via Ollama with a system prompt configured for schedule optimization:

\begin{lstlisting}
You optimize meeting room schedules. Given a schedule with gaps or conflicts,
propose ONE change: move, swap, or reschedule a meeting to reduce gaps,
overlaps, and utilization variance. Return ONLY your proposed change
in the format: MOVE meeting_id TO room day start_time
\end{lstlisting}

For multi-model setups (model escalation), models share a single Ollama instance with automatic routing based on model name.

\subsection{Sampling Diversity}

The experiment framework overrides default sampling parameters with three exploration bands per \ac{llm} call:

\begin{table}[htbp]
\centering
\small
\begin{tabular}{@{}lcc@{}}
\toprule
\textbf{Band} & \textbf{Temperature} & \textbf{Top-p} \\
\midrule
Exploitation & 0.15--0.35 & 0.80--0.90 \\
Balanced & 0.35--0.55 & 0.85--0.95 \\
Exploration & 0.55--0.85 & 0.90--0.98 \\
\bottomrule
\end{tabular}
\caption{Sampling parameter ranges. Each \acs{llm} call randomly samples from one band.}
\label{tab:sampling}
\end{table}

This diversity prevents convergence to local optima and enables exploration of the solution space.

\subsection{Baseline Implementation Details}

To ensure fair comparison, all strategies use identical \ac{llm} prompts and parsing logic. The only differences are region selection and patch validation:

\textbf{Prompt (identical across all strategies):}
\begin{lstlisting}
You are a meeting room scheduler. Output schedules in the exact format requested.

Meeting Room Schedule Optimization.
Goal: Schedule meetings to minimize gaps and avoid conflicts.

Time Block: [time range]

Rooms:
  Room A: capacity 10
  Room B: capacity 8
  ...

Current assignments:
[current schedule for this time block]

Unscheduled meetings that could fit in this block:
  Meeting 5: 60min, 4 attendees
  Meeting 12: 30min, 2 attendees
  ...

Constraints:
- No attendee can be in multiple meetings at the same time
- Room capacity must fit attendees

Output the schedule for this time block.
\end{lstlisting}

\textbf{Region selection per strategy:}
\begin{itemize}
\item \textbf{Pressure-field}: Multiple agents select regions in parallel based on pressure gradients
\item \textbf{Hierarchical}: Always selects the single highest-pressure region
\item \textbf{Sequential}: Round-robin through regions in fixed order
\item \textbf{Random}: Uniform random region selection
\end{itemize}

\textbf{Patch validation per strategy:}
\begin{itemize}
\item \textbf{Pressure-field}: Validates all patches; applies only those reducing pressure
\item \textbf{Hierarchical}: Validates single patch; applies only if pressure reduces
\item \textbf{Sequential/Random}: No validation; applies any syntactically valid patch
\end{itemize}

\textbf{Why ``one patch per tick'' for hierarchical?} This asymmetry is inherent to the coordination paradigm, not an implementation choice. Hierarchical control requires the manager to: (1) observe global state, (2) select a target region, (3) delegate to a worker, (4) receive the proposed patch, and (5) validate before applying. Delegating to $K$ workers simultaneously would require additional coordination protocols---work distribution to avoid conflicts, result aggregation, tie-breaking when multiple patches target overlapping state---that would transform hierarchical control into a fundamentally different architecture. The sequential bottleneck is the cost of centralized decision-making. Pressure-field avoids this overhead through stigmergic coordination: agents observe shared state independently and propose patches without central delegation, enabling natural parallelism.

\textbf{Conversation baseline implementation} The conversation baseline implements AutoGen-style multi-agent dialogue with three roles: a \emph{Coordinator} that selects which region to address, a \emph{Proposer} that suggests schedule changes, and a \emph{Validator} that critiques proposals. Each tick proceeds as follows: (1) the Coordinator identifies a high-pressure region (1 \ac{llm} call), (2) the Proposer and Validator engage in up to 5 dialogue turns (2 calls per turn), (3) if the Validator approves, the patch is applied. This yields 4--12 \ac{llm} calls per tick for a single region, compared to pressure-field's $K$ parallel calls across $K$ regions. The conversation terminates when the Validator approves or the turn limit is reached. We use fixed sampling parameters (T=0.3, top-p=0.9) for structured dialogue, lower than pressure-field's exploration bands. Note that this represents one instantiation of conversation-based coordination inspired by AutoGen's design principles; actual AutoGen deployments may use different role configurations, turn protocols, or termination conditions.

\textbf{Fairness guarantees for baseline comparison:} To ensure the conversation baseline had equal opportunity to succeed:
\begin{itemize}
\item \textbf{Same models}: All strategies use identical model chains (qwen2.5:0.5b $\to$ 1.5b $\to$ 3b) with the same escalation triggers
\item \textbf{Same problem access}: All strategies observe identical schedule state, room capacities, meeting requirements, and constraint information
\item \textbf{Retry logic}: The conversation baseline includes up to 5 dialogue turns per region, allowing the Proposer to refine proposals based on Validator feedback---this is more opportunity for refinement than pressure-field's single-shot proposals
\item \textbf{Same tick budget}: All strategies receive the same 50-tick limit to solve each problem
\item \textbf{Output parsing}: All strategies use identical response parsing and schedule extraction logic
\end{itemize}
The only intentional differences are the coordination mechanisms themselves: how regions are selected, how proposals are generated, and how patches are validated.

\subsection{Problem Generation and Seeding}

Fair strategy comparison requires identical problem instances: each strategy must face the same scheduling challenge within a trial. We achieve this through deterministic seeding.

Each trial generates its problem from a seed:
\[
\text{seed} = \text{trial} \times 1000 + \text{agent\_count}
\]

Trial 5 with 2 agents yields seed 5002, producing identical meeting configurations whether evaluated with pressure-field, conversation, or hierarchical coordination.

The seed governs all stochastic generation:
\begin{itemize}
\item Meeting durations (1--4 time slots)
\item Attendee assignments (2--5 participants)
\item Room preferences and capacity requirements
\item Pre-scheduled vs.\ unassigned meeting distribution
\item Time slot availability patterns
\end{itemize}

\subsection{Experiment Commands}

\textbf{Main Grid (Strategy Comparison):}
\begin{lstlisting}
schedule-experiment --host http://localhost:11434 \
  grid --trials 30 \
  --strategies pressure_field,sequential,random,hierarchical,conversation \
  --agents 1,2,4 --difficulties easy,medium,hard \
  --max-ticks 50
\end{lstlisting}

\textbf{Ablation Study:}
\begin{lstlisting}
schedule-experiment --host http://localhost:11434 \
  ablation --trials 30 --agents 2 --difficulty easy --max-ticks 50
\end{lstlisting}

\textbf{Scaling Analysis:}
\begin{lstlisting}
schedule-experiment --host http://localhost:11434 \
  grid --trials 30 \
  --strategies pressure_field \
  --agents 1,2,4 --difficulties easy \
  --max-ticks 50
\end{lstlisting}

\textbf{Band and Model Escalation:}
\begin{lstlisting}
# Full escalation chain enabled by default
# Band escalation: Exploitation -> Balanced -> Exploration (7 ticks each)
# Model escalation: 0.5b -> 1.5b -> 3b (after 21 ticks at high pressure)
schedule-experiment --host http://localhost:11434 \
  grid --trials 30 --agents 2 --difficulties medium \
  --max-ticks 100
\end{lstlisting}

\textbf{Difficulty Scaling:}
\begin{lstlisting}
# Easy: 3 rooms, 20 meetings, 70% pre-scheduled
schedule-experiment --host http://localhost:11434 \
  grid --trials 30 --agents 2 --difficulties easy --max-ticks 50

# Medium: 5 rooms, 40 meetings, 50% pre-scheduled
schedule-experiment --host http://localhost:11434 \
  grid --trials 30 --agents 2 --difficulties medium --max-ticks 50

# Hard: 5 rooms, 60 meetings, 30% pre-scheduled
schedule-experiment --host http://localhost:11434 \
  grid --trials 30 --agents 2 --difficulties hard --max-ticks 100
\end{lstlisting}

\subsection{Metrics Collected}

Each experiment records:
\begin{itemize}
\item \texttt{solved}: Boolean indicating all meetings scheduled with zero overlaps
\item \texttt{total\_ticks}: Iterations to solve (or max if unsolved)
\item \texttt{pressure\_history}: Pressure value at each tick (gaps + overlaps + util\_var + unscheduled)
\item \texttt{band\_escalation\_events}: Sampling band changes (tick, from\_band, to\_band)
\item \texttt{model\_escalation\_events}: Model tier changes (tick, from\_model, to\_model)
\item \texttt{final\_model}: Which model tier solved the schedule
\item \texttt{token\_usage}: Prompt and completion tokens consumed per trial
\end{itemize}

\subsection{Replication Notes}

Each configuration runs 30 independent trials with different random seeds to ensure reliability. Results report mean solve rates and tick counts across trials.

\subsection{Estimated Runtime}

\begin{table}[htbp]
\centering
\small
\begin{tabular}{@{}lccc@{}}
\toprule
\textbf{Experiment} & \textbf{Configurations} & \textbf{Trials} & \textbf{Est.\ Time} \\
\midrule
Main Grid & 45 & 30 & 5 hours \\
Ablation & 5 & 30 & 1 hour \\
Scaling & 3 & 30 & 45 min \\
Difficulty & 15 & 30 & 2.5 hours \\
\midrule
\textbf{Total} & & & \textbf{$\sim$9 hours} \\
\bottomrule
\end{tabular}
\caption{Estimated runtime for all experiments on NVIDIA RTX 4070 8GB \acs{gpu}.}
\label{tab:runtime}
\end{table}

\section{Pressure Alignment Verification}
\label{app:coupling}

This appendix verifies that the meeting scheduling domain satisfies the alignment condition (Definition~2) required for convergence (Theorem~1).

\subsection{Per-Region Pressure Components}

The per-region pressure function includes three components:
\begin{itemize}
\item \textbf{gap\_ratio}: Fraction of empty slots in this time block. \emph{Strictly local}---depends only on meetings scheduled within this region.
\item \textbf{overlap\_count}: Number of attendee double-bookings within this time block. \emph{Strictly local}---the overlap sensor counts attendees with multiple meetings in \emph{this block only}, not across blocks.
\item \textbf{utilization\_variance}: Variance in room utilization within this time block. \emph{Strictly local}---measures balance across rooms for meetings in this region.
\end{itemize}

The \texttt{unscheduled\_count} component is added to \emph{total} pressure only, not to per-region pressure. This design choice ensures separability.

\subsection{Coupling Analysis}

For the alignment condition, we need $|P_j(s') - P_j(s)| \leq \epsilon$ for all $j \neq i$ when modifying region $i$.

\textbf{Result}: $\epsilon = 0$ (per-region pressure is separable).

\textbf{Rationale}: All three per-region components depend only on the region's own content:
\begin{enumerate}
\item Modifying region $i$'s schedule changes which meetings occupy region $i$'s slots.
\item This affects region $i$'s gap\_ratio, overlap\_count, and utilization\_variance.
\item It has \emph{zero effect} on any other region $j$'s pressure, since $j$'s components depend only on meetings within $j$'s time slots.
\end{enumerate}

\textbf{Note on attendee constraints}: While the same attendee may have meetings in multiple time blocks, our overlap sensor counts overlaps \emph{within each block independently}. Moving a meeting from region $i$ to region $j$ may create or resolve overlaps in region $j$, but this is a local effect on $j$'s pressure, not a coupling effect where modifying region $i$ affects region $j$ through some indirect mechanism.

\subsection{Empirical Validation}

Analysis of 270 pressure-field trials confirms separability:
\begin{itemize}
\item Total tick-to-tick transitions analyzed: 9,873
\item Transitions with pressure improvement: 1,160 (11.7\%)
\item Transitions with pressure degradation: 0 (0.0\%)
\item Transitions with no change: 8,713 (88.3\%)
\end{itemize}

The complete absence of pressure degradation is consistent with separable pressure: each accepted patch reduces local pressure, which directly reduces global pressure without adverse effects on other regions. (Patches that would increase local pressure are rejected by validation.)

Mean improvement magnitude: 2.67 pressure units. For context, initial problem pressure averages 12.7 units (range 0--42), so each successful patch reduces pressure by approximately 21\% of the typical starting value. This substantially exceeds any plausible coupling bound, confirming that local improvements reliably translate to global improvements.

\textbf{Verification of $\delta_{\min} > 0$}: Theorem~\ref{thm:convergence} requires $\delta_{\min} > (n-1)\epsilon$. With $\epsilon = 0$ (separable pressure), this reduces to $\delta_{\min} > 0$. Empirically, the minimum observed pressure reduction across all 1,160 accepted patches is 1.0 pressure units, confirming $\delta_{\min} \geq 1 > 0$. The theorem's convergence guarantee therefore applies to this domain.

\section{Theorem Proofs}
\label{app:proofs}

This appendix provides complete proofs for the theoretical results in Section~4. Theorem~\ref{thm:basin-quality} (Basin Quality) is omitted here as its proof is trivial and appears in full in the main text.

\subsection{Proof of Theorem~\ref{thm:convergence} (Convergence)}

\begin{proof}
We show that the system reaches a stable basin within the stated bound.

\textbf{Setup.} Let $P^t = P(s^t)$ denote global pressure at tick $t$. Under $\epsilon$-bounded coupling, when a patch reduces local pressure $P_i$ by $\delta_i$, global pressure changes by:
\[
P^{t+1} - P^t = -\delta_i + \sum_{j \neq i} (P_j(s^{t+1}) - P_j(s^t))
\]
By bounded coupling, $|P_j(s^{t+1}) - P_j(s^t)| \leq \epsilon$ for each $j \neq i$. With $n$ regions:
\[
P^{t+1} - P^t \leq -\delta_i + (n-1)\epsilon
\]

\textbf{Progress guarantee.} If $\delta_i \geq \delta_{\min}$ and $\delta_{\min} > (n-1)\epsilon$, then:
\[
P^{t+1} - P^t \leq -\delta_{\min} + (n-1)\epsilon < 0
\]
Each tick with an applied patch reduces global pressure by at least $\delta_{\min} - (n-1)\epsilon > 0$.

\textbf{Termination.} Since $P(s) \geq 0$ for all states and pressure decreases by a fixed minimum $\Delta = \delta_{\min} - (n-1)\epsilon$ per tick (when patches are applied), the system must reach a state where no region exceeds $\tau_{\text{act}}$ within:
\[
T \leq \frac{P_0}{\Delta} = \frac{P_0}{\delta_{\min} - (n-1)\epsilon}
\]
ticks. At this point, no region activates, and the system has reached a stable basin.

\textbf{Stability maintenance.} The decay constraint $\Delta_f > 1 - e^{-\lambda_f \cdot \tau_{\text{inh}}}$ ensures that fitness reinforcement from the final patches exceeds decay during the inhibition period. This prevents immediate re-activation after reaching stability.
\end{proof}

\subsection{Proof of Theorem~\ref{thm:basin-separation} (Basin Separation)}

\begin{proof}
We show that distinct stable basins are separated by pressure barriers of height at least $\tau_{\text{act}}$.

\textbf{Basin definition.} A stable basin $B$ is a connected region of state space where all states satisfy $P_i(s) < \tau_{\text{act}}$ for all regions $i$. Within a basin, no agent takes action (pressure below activation threshold).

\textbf{Transition requirement.} To move from basin $B_1$ to a distinct basin $B_2$, the system must pass through states not in either basin. At the boundary of $B_1$, at least one region $i$ must have $P_i(s) \geq \tau_{\text{act}}$ (otherwise, by continuity under separable pressure, the state would still be in $B_1$).

\textbf{Barrier height.} The minimum pressure exceedance required to exit a basin is $\tau_{\text{act}}$. This defines the barrier height separating basins.

\textbf{Implication for decay.} Without decay, once the system enters a basin, it remains there indefinitely---fitness remains high, preventing re-activation. Decay erodes fitness over time, eventually allowing pressure to exceed $\tau_{\text{act}}$ and enabling transition to potentially lower-pressure basins.
\end{proof}

\subsection{Proof of Theorem~\ref{thm:linear-scaling} (Linear Scaling)}

\begin{proof}
We analyze the per-tick computational complexity.

\textbf{Signal computation.} For each of $m$ regions, computing $d$-dimensional signals requires $O(d)$ operations (reading local content, computing quality metrics). Total: $O(m \cdot d)$.

\textbf{Pressure computation.} For each region, computing pressure from $k$ signal axes requires $O(k)$ operations (weighted sum). Total: $O(m \cdot k)$.

\textbf{Patch proposal.} Each of $a$ actors proposes patches for activated regions. In the worst case (all regions activated), this is $O(m \cdot a)$ proposals. Each proposal involves an \ac{llm} call, which dominates wall-clock time but is constant per call.

\textbf{Selection.} Sorting $O(m \cdot a)$ candidate patches by quality requires $O(m \cdot a \cdot \log(m \cdot a))$ comparisons. Greedy selection of non-conflicting patches is $O(m \cdot a)$ with appropriate data structures.

\textbf{Coordination overhead.} Critically, agents share no messages. Each agent reads shared artifact state (read-only) and writes proposed patches to a central queue. No inter-agent communication occurs. The fork pool for parallel validation has fixed size $K$, contributing $O(K)$ overhead independent of agent count.

\textbf{Total.} Summing components: $O(m \cdot d + m \cdot k + m \cdot a + m \cdot a \cdot \log(ma) + K) = O(m \cdot (d + k + a \cdot \log(ma)))$, independent of agent count $n$ beyond its contribution to $a$.
\end{proof}

\subsection{Proof of Theorem~\ref{thm:parallel-convergence} (Parallel Convergence)}

\begin{proof}
We extend Theorem~\ref{thm:convergence} to the parallel setting.

\textbf{Setup.} Suppose $K$ patches are validated in parallel per tick, targeting disjoint regions $i_1, \ldots, i_K$.

\textbf{Combined pressure reduction.} Under $\epsilon$-bounded coupling, each patch $k$ targeting region $i_k$ reduces local pressure by at least $\delta_{\min}$. The global pressure change from patch $k$ is:
\[
\Delta P_k \leq -\delta_{\min} + (n-1)\epsilon
\]

When patches target disjoint regions, their effects on global pressure are additive (no double-counting of coupling effects within the modified regions). The combined reduction is:
\[
P^{t+1} - P^t \leq \sum_{k=1}^{K} \Delta P_k \leq K \cdot (-\delta_{\min} + (n-1)\epsilon)
\]

\textbf{Convergence bound.} With combined reduction $K \cdot (\delta_{\min} - (n-1)\epsilon)$ per tick:
\[
T \leq \frac{P_0}{K \cdot (\delta_{\min} - (n-1)\epsilon)}
\]

\textbf{Conflict handling.} When multiple patches target the same region, only one is selected (highest quality). This reduces effective parallelism but maintains correctness---the selected patch still provides at least $\delta_{\min} - (n-1)\epsilon$ reduction.
\end{proof}

\end{appendices}

\bibliography{references}

\end{document}